\documentclass[acmsmall]{acmart}

\AtBeginDocument{%
  \providecommand\BibTeX{{%
    \normalfont B\kern-0.5em{\scshape i\kern-0.25em b}\kern-0.8em\TeX}}}

\usepackage[utf8]{inputenc} % allow utf-8 input
\usepackage{hyperref}       % hyperlinks
\usepackage{url}            % simple URL typesetting
\usepackage{booktabs}       % professional-quality tables
\usepackage{amsfonts}       % blackboard math symbols
\usepackage{nicefrac}       % compact symbols for 1/2, etc.
\usepackage{microtype}      % microtypography

\usepackage{amsmath,amsfonts}
\usepackage{algorithmic}
\usepackage{graphicx}
\usepackage{textcomp}
\usepackage{xcolor}
\usepackage{multirow}
\usepackage{algorithm}
\usepackage{bm}
\usepackage{subfigure}
\usepackage{caption}
\usepackage{amsthm}
\usepackage{afterpage}
\usepackage{longtable}
\usepackage{adjustbox}

\setcopyright{acmlicensed}
\acmJournal{TOIS}
\acmYear{2021} \acmVolume{1} \acmNumber{1} \acmArticle{1} \acmMonth{1} \acmPrice{15.00}\acmDOI{XXXXXXX.XXXXXXX}

\begin{document}

%%
%% The "title" command has an optional parameter,
%% allowing the author to define a "short title" to be used in page headers.
\title{AutoML for Deep Recommender Systems: A
Survey}

%%
%% The "author" command and its associated commands are used to define
%% the authors and their affiliations.
%% Of note is the shared affiliation of the first two authors, and the
%% "authornote" and "authornotemark" commands
%% used to denote shared contribution to the research.
\author{Ruiqi Zheng}
\authornote{Both authors contributed equally to this research.}
\affiliation{%
  \institution{The University of Queensland}
  \city{Brisbane}
  \state{Queensland}
  \country{Australia}
  \postcode{4072}
}
\email{ruiqi.zheng@uq.net.au}
\orcid{0000-0001-5831-1838}

% \affiliation{%
%   \institution{The University of Queensland}
%   \city{Brisbane}
%   \state{Queensland}
%   \country{Australia}
%   \postcode{4072}
% }
% \affiliation{%
%   \institution{Southern University of Science and Technology}
%   \city{Shenzhen}
%   \country{China}
% }

\author{Liang Qu}
\authornotemark[1]
\affiliation{%
  \institution{The University of Queensland}
  \city{Brisbane}
  \state{Queensland}
  \country{Australia}
  \postcode{4072}
}
\email{l.qu1@uq.net.au}
\orcid{0000-0002-2755-7592}

\author{Bin Cui}
\affiliation{%
  \institution{Peking University}
  \streetaddress{5 Yiheyuan Rd, Haidian District}
  \city{Beijing}
  \country{China}}
\email{bin.cui@pku.edu.cn}
\orcid{0000-0003-1681-4677}

\author{Yuhui Shi}
\authornote{Corresponding authors.}
\affiliation{%
  \institution{Southern University of Science and Technology}
  \city{Shenzhen}
  \country{China}
}
\email{shiyh@sustech.edu.cn}
\orcid{0000-0002-8840-723X}

\author{Hongzhi Yin}
\authornotemark[2]
\affiliation{%
 \institution{University of Queensland}
  \streetaddress{St Lucia QLD 4072}
  \city{Brisbane}
  \state{Queensland}
  \country{Australia}
  \postcode{4072}}
\email{h.yin1@uq.edu.au}
\orcid{0000-0003-1395-261X}

\thanks{This work was supported by the Australian Research Council Future Fellowship (Grant No. FT210100624), the Discovery Project (Grant No. DP190101985), the National Natural Science Foundation
of China (Grant No. 61761136008), the Shenzhen Fundamental Research Program (Grant No. JCYJ20200109141235597), the Guangdong Basic and Applied Basic Research Foundation (Grant No. 2021A1515110024), the Shenzhen Peacock Plan (Grant No. KQTD2016112514355531), the Program
for Guangdong Introducing Innovative and Entrepreneurial Teams (Grant No. 2017ZT07X386).}

%%
%% By default, the full list of authors will be used in the page
%% headers. Often, this list is too long, and will overlap
%% other information printed in the page headers. This command allows
%% the author to define a more concise list
%% of authors' names for this purpose.
\renewcommand{\shortauthors}{Ruiqi Zheng, et al.}

%%
%% The abstract is a short summary of the work to be presented in the
%% article.
\begin{abstract}
Recommender systems play a significant role in information filtering and have been utilized in different scenarios, such as e-commerce and social media. With the prosperity of deep learning, deep recommender systems show superior performance by capturing non-linear information and item-user relationships. However, the design of deep recommender systems heavily relies on human experiences and expert knowledge. To tackle this problem, Automated Machine Learning (AutoML) is introduced to automatically search for the proper candidates for different parts of deep recommender systems. This survey performs a  comprehensive review of the literature in this field. Firstly, we propose an abstract concept for AutoML for deep recommender systems (AutoRecSys) that describes its building blocks and distinguishes it from conventional AutoML techniques and recommender systems. Secondly, we present a taxonomy as a classification framework containing feature selection search, embedding dimension search, feature interaction search, model architecture search, and other components search. Furthermore, we put a particular emphasis on the search space and search strategy, as they are the common thread to connect all methods within each category and enable practitioners to analyze and compare various approaches. Finally, we propose four future promising research directions that will lead this line of research.
\end{abstract}

%%
%% The code below is generated by the tool at http://dl.acm.org/ccs.cfm.
%% Please copy and paste the code instead of the example below.
%%


%%
%% Keywords. The author(s) should pick words that accurately describe
%% the work being presented. Separate the keywords with commas.

%%
%% This command processes the author and affiliation and title
%% information and builds the first part of the formatted document.
\maketitle

\section{Introduction}
\label{intro}
The amount of information has increased tremendously due to the fast expansion of the internet. Users find it challenging to choose what interests them among many options due to the abundance of information. Recommender systems \cite{lu2012recommender,zhang2019deep} have been utilized in different scenarios, such as e-commerce \cite{zhou2018micro,li2015online} and social media \cite{xu2015social,covington2016deep}, to improve the user experience. Users count on recommender systems to help them deal with information overload problems and find what they are interested in among the immense sea of options. An effective recommender system predicts users' preferences based on users' previous engagements \cite{Chen2022Automated,lin2022adafs,wang2022autofield}.

Over the last several years, the primary model framework of recommender systems has been developed from neighborhood techniques \cite{bell2007scalable,linden2003amazon,sarwar2001item} to representation learning \cite{covington2016deep,koren2015advances,koren2009matrix,sedhain2015autorec,wang2015collaborative}. Item-based neighborhood approaches \cite{bell2007scalable,linden2003amazon,sarwar2001item} proactively recommend items that are similar to consumers' previous interacted items. Neighborhood techniques have been proven effective in real-world applications due to interpretability and simplicity. In comparison, representation-based methods represent the users and items in the latent embedding space. As the most classic representation-based methods, matrix factorization methods \cite{koren2015advances,koren2009matrix} are designed to handle the data sparsity problem with dimensionality reduction. 

With the prosperity of deep learning \cite{lecun2015deep}, deep neural networks (DNN) generate more complicated and informative representations. Theoretically, one single-layer perceptron can mimic any function with enough computation source, and data \cite{lu2017expressive}. Deep recommender systems that integrate deep learning techniques into recommender systems have been proposed to capture the non-linear information and item-user relationships \cite{okura2017embedding,9101653}. Therefore, they have become favored in the industrial and academic world.

\begin{figure*} 

\centering
\includegraphics[width=\textwidth]{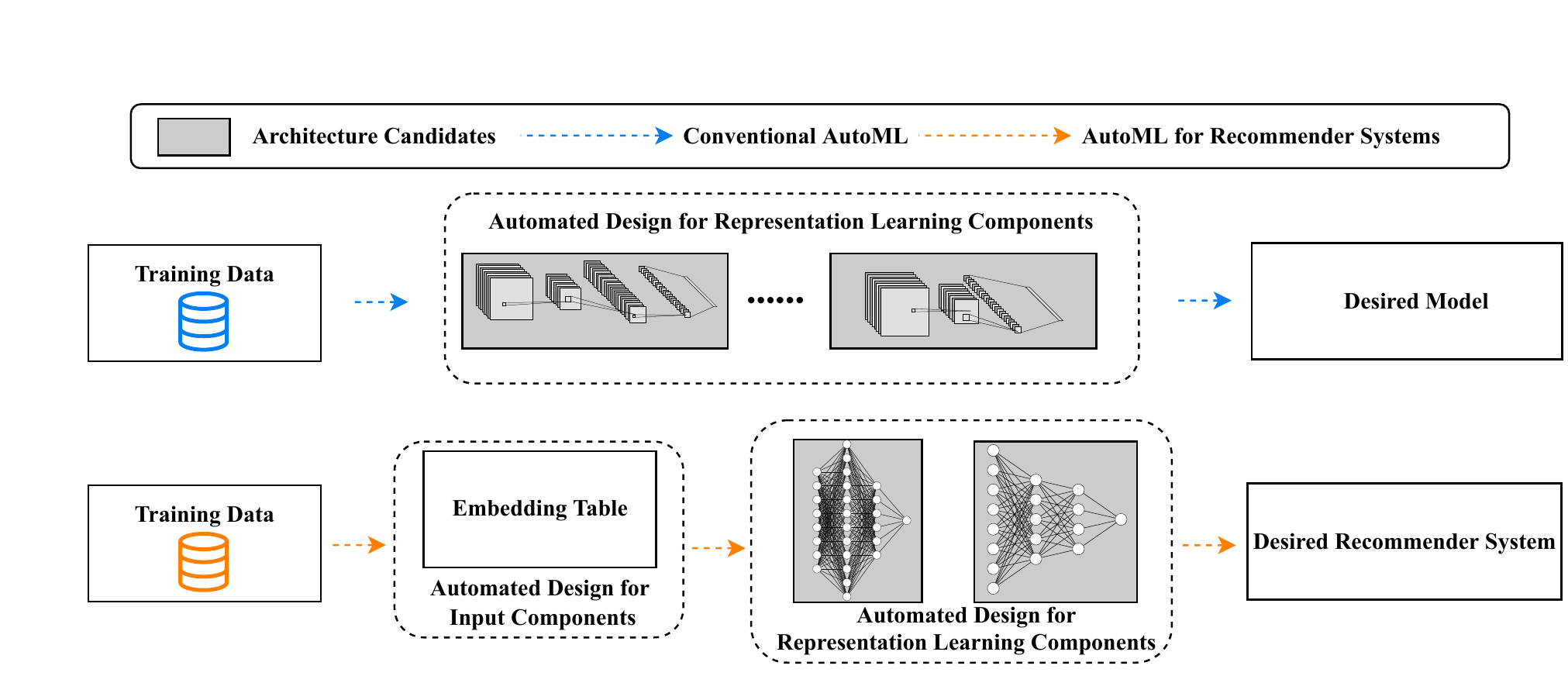}
\caption{Illustration of AutoML for recommender systems and conventional AutoML.}
\label{fig:conventional}
\end{figure*}

Deep recommender systems \cite{cheng2016wide,covington2016deep} typically have four components. The input layer generates the binary features from raw data. The embedding layer maps the binary features into low-dimensional feature space. The interaction layer finds the powerful feature interaction that benefits the model's performance, and the prediction layer generates the model's prediction. Section 2 will introduce the mathematical form of these four components in detail. 

Although deep recommender systems show promising and encouraging results, they heavily require human experiences, and the lack of careful design for different components leads to sub-optimal performance. For example, in the embedding layer, most existing methods \cite{guo2017deepfm,covington2016deep} simply assign a uniform embedding dimension for all features, which suffers from the issues such as resource consumption, computation cost, and model representation ability. In the interaction layer, all the $2^{nd}$ order feature interactions are calculated \cite{rendle2010factorization,qu2016product,cheng2016wide,guo2017deepfm}, which introduces excessive noise to the model and complicates the training procedure. Automatically designed methods for different components of the deep recommender systems are urgently needed to alleviate humans from complicated and time-consuming work. 

Recently, Automated Machine Learning (AutoML) \cite{yao2018taking} has emerged as a promising way to automate some components or the whole machine learning pipeline. Compared with conventional recommender systems, which require experts to develop a specific model, AutoML for deep recommender systems (AutoRecSys) outputs the well-performed deep recommender systems in a data-oriented and task-specific manner by automatically designing different opponents and alleviating human effort. It is more capable of discovering a well-performed model when encountering various application scenarios and outperforming traditional methods. It focuses on the challenges brought by the design of compact search space and efficient search strategy rather than developing one single recommender system model.

As shown in Fig.~\ref{fig:conventional}, AutoML automatically designs the representation components, such as pooling, convolution, and the number of layers, in computer version applications \cite{saikia2019autodispnet,yang2021medmnist}. However, AutoRecSys is not simply an application of AutoML techniques but is faced with unique challenges \cite{cheng2020differentiable}. Most existing AutoML methods primarily concentrate on the automatic design of representation learning components, and input components have received little attention because the majority of research is conducted on image understanding issues \cite{liu2018darts,pmlr-v80-pham18a,xie2018snas,zoph2018learning}, where the pixels of the image as the input component do not require creating features from the data since they are already in floating-point form. However, for deep recommender systems, the input component like the embedding matrix is the primary factor of memory consumption \cite{wang2020next} in comparison with other parameters such as biases and weights. How to properly learn the features from the raw data dramatically influences other components and is crucial to the final model performance. AutoML does not reveal universal or principled approaches to learning features from data and only makes limited progress in this direction \cite{yao2018taking}.

In industry, AutoRecSys has been deployed in large-scale real-world applications to provide discriminative and informative recommendation results. For example, Huawei Noah's Ark Lab implements AutoFIS \cite{liu2020autofis} to automatically search for beneficial feature interactions \cite{liu2020autofis} and illustrates the significant improvements in the Huawei App Store recommender task by a 10-day online A/B test.

\begin{figure*} 
\centering
\includegraphics[width=\textwidth]{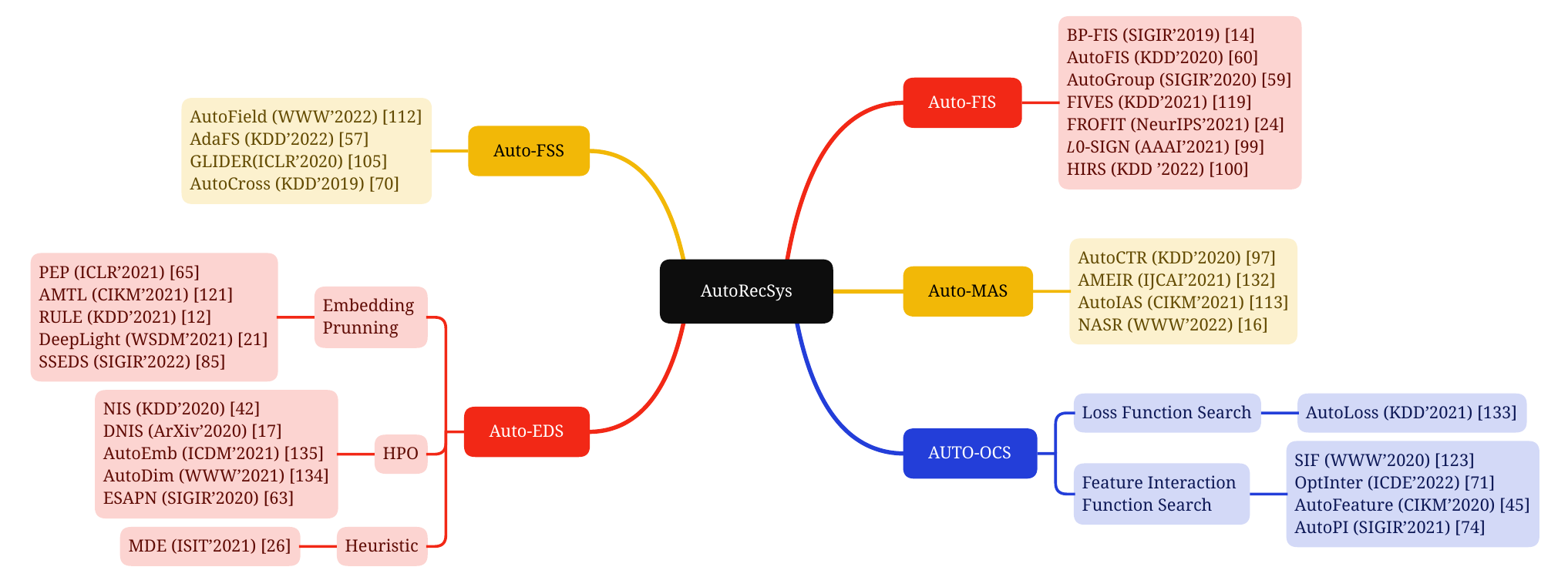}
\caption{Categorization for AutoRecSys methods.}
\label{fig:mind}
\end{figure*}

Given the significant growth rate of AutoRecSys, we believe it is essential to synthesize and describe representative techniques within a uniform and comprehensible paradigm. To the extent of our knowledge, the most relevant survey about automated machine learning for deep recommender systems published formally is a short paper \cite{Chen2022Automated}. Our work has the following difference from the above one: (1) Our survey includes more representative AutoRecSys methods from top venues, including MDE (ISIT'2021), SSEDS (SIGIR'2022), $L_{0}$-SIGN (AAAI'2021), HIRS (KDD'2022), NASR (WWW'2022), OptInter (ICDE'2022). (2) Our work is the first survey to comprehensively review AutoRecSys and present a taxonomy, which appeared on Arxiv on March 25, 2022. (3) Our work includes search space complexity and experiments, which horizontally compare AutoRecSys methods mathematically and empirically. (4) We summarize the core steps of AutoRecSys, and elaborate our analysis on the strengths and defects of AutoRecSys methods instead of generally introducing every model.

The contributions of this survey paper are threefold: 
\begin{itemize}
\item We propose an abstract concept \textbf{AutoML for Deep Recommender Systems (AutoRecSys)} that clarifies its procedures and differences from conventional AutoML and conventional recommender systems. It states that AutoML for deep recommender systems outputs the well-performed deep recommender systems in a data-oriented and task-specific manner by automatically designing different opponents and alleviating human effort. To our best knowledge, this is the first survey that proposes the abstract concept and systematically reviews the literature of AutoRecSys.

\item The second contribution is the introduction of a taxonomy that classifies AutoML methods for recommendation systems. It contains feature selection search, embedding dimension search, feature interaction search, model architecture search, and other components search, as shown in Fig.~\ref{fig:mind}. Moreover, we put a specific emphasis on the search space and search strategy, as they are the common thread to connect all methods within each category and enable practitioners to analyze and compare various approaches. 

\item We state our own opinions on
existing works and discuss their potential drawbacks. Furthermore, we propose four future promising research directions leading this line of research.
 
\end{itemize}

This survey paper aims to equip potential new users of AutoML for deep recommender systems with proven and practical techniques. As we plan to survey a broad range of techniques in this field, we cannot cover every methodical detail, and we do not claim to include all available research. Instead, we tend to analyze and summarize common grounds as well as the diversity of approaches. Thus, the prevailing research directions in AutoRecSys can be outlined.

The rest of the paper is organized as follows. Section~\ref{sec:classification} describes how we categorized the approaches. Section~\ref{sec:background} introduces the background of deep recommender systems and frequently-used skills in AutoML for deep recommender systems inspired by Neural Architecture Search (NAS). The five categories in taxonomy: automated feature selection search, automated embedding dimension search, automated feature interaction search, automated model architecture search, and automated other components search, are presented from Section~\ref{sec:autofss} to Section~\ref{sec:autoocs}. In Section~\ref{sec:experiment}, horizontal comparison and empirical analysis for AutoRecSys are performed. In Section~\ref{sec:future}, future directions are discussed, followed by the conclusions in Section~\ref{sec:conclusion}.

\section{Classification of Approaches}
\label{sec:classification}
\begin{figure} 
\centering

\includegraphics[scale=0.68]{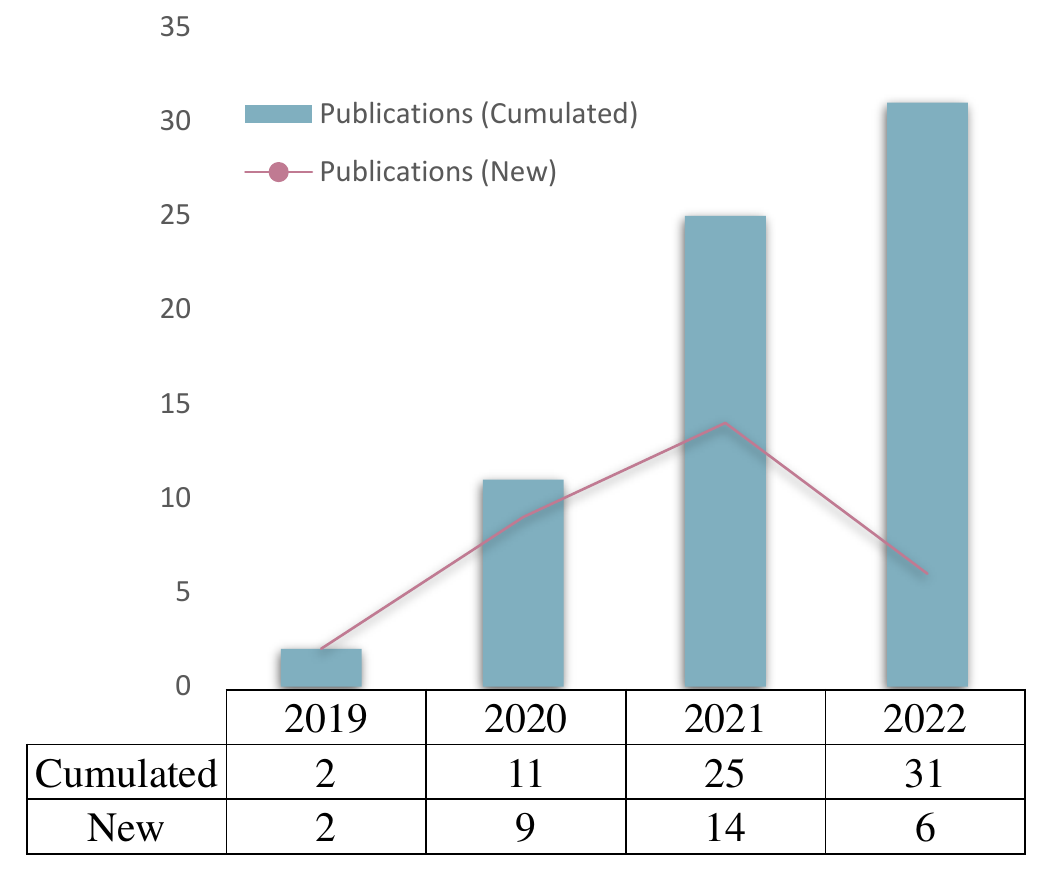}

\caption{Annual publications in the field of AutoRecSys. Numbers are based on our survey. We conducted the survey in October 2022. Articles presented in late 2022 most likely had not been published and thus were not discovered through our search.}
\label{fig:count}
\end{figure}

To understand how the concept of AutoML for deep recommender systems is implemented, we developed a comprehensive classification of existing methodologies. We do not claim to include all available research since our goals are to analyze different methods and determine their similarities or distinctions. The representative methods are selected from top computer science journals or conferences, such as KDD and WWW. The count of papers published per year is illustrated in Fig.~\ref{fig:count}. In this section, we describe the analysis questions, which determine our classification, and then our literature surveying procedure.

\subsection{Analysis Questions}
Our guiding question is how can the different components of the deep recommender systems be automatically designed to fit various scenarios and data. Under the guiding question, our survey paper focuses on the following three questions. (1) Which components of the models are automatically designed? (2) How is the compact search space designed so that it is general enough to include popular human-crafted models and not too general to prevent the success of the search for the new models? (3) How is the search algorithm designed so that exploration and exploitation can be balanced to enhance search efficiency and effectiveness?

\subsection{Literature Surveying Procedure}
To comprehensively answer the above analysis questions, we reviewed a wide range of papers discussing AutoML for deep recommender systems. A comparative and iterative literature review process is implemented. In the first round, a set of publications is inspected and summarized based on their answers to our analysis questions. We found that many papers automatically design the same component and encounter similar challenges. Therefore we classified those papers in the same category. The second round focuses on similar challenges within the same category and analyses the proposed methods to deal with similar challenges. In the third cycle, we sorted publications and enlarged our set of publications. Consequently, there is an extensive literature review where a distilled taxonomy includes various papers.

\begin{figure*} 
\centering
\includegraphics[width=\textwidth]{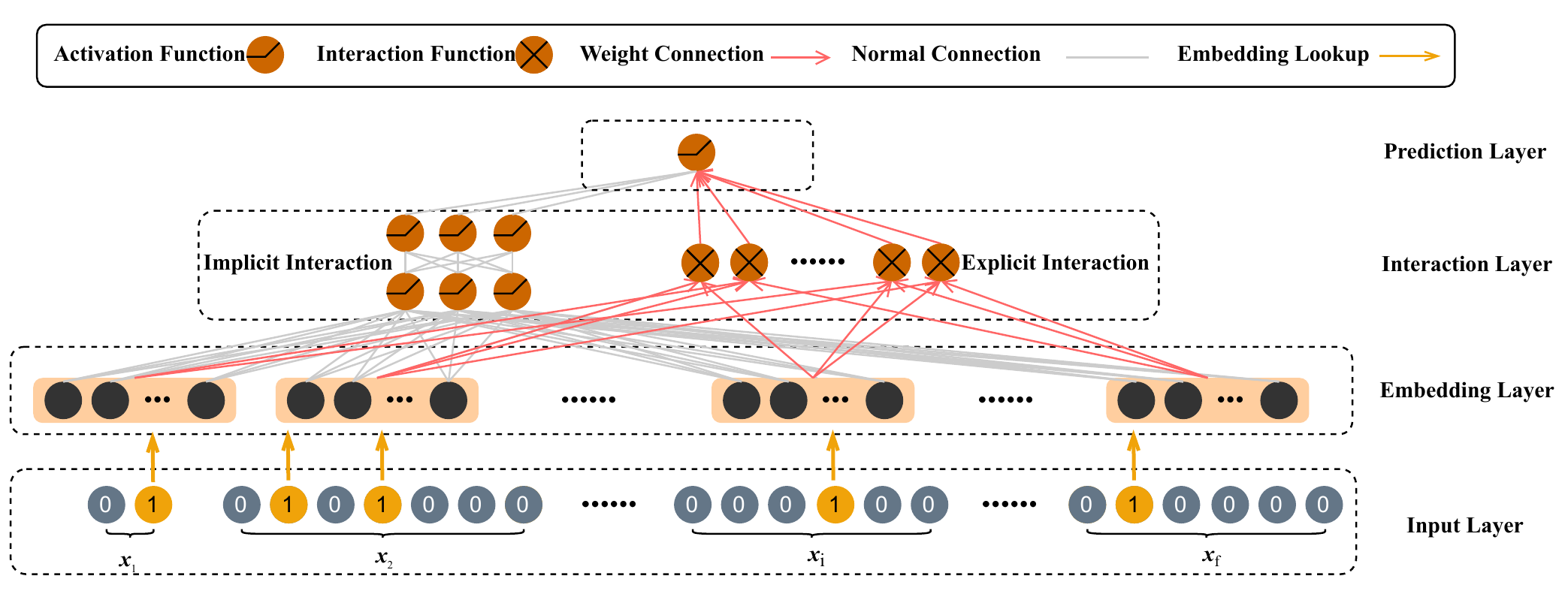}
\caption{Illustration of deep recommender systems.}
\label{fig:drs}
\end{figure*}

\section{Backgrounds}
\label{sec:background}
\subsection{Deep Recommender Systems}

The frequently used notations are listed in Table~\ref{table:sign}. Deep learning has been widely applied to recommender systems because it captures the non-linear information and item-user relationships. Deep recommender systems \cite{meng2021general} typically have four layers, as shown in Fig. \ref{fig:drs}: input layer, embedding layer, interaction layer, and prediction layer. We will give an introduction to the above components.

\subsubsection{Input Layer}
 The input data usually contains three types of information, namely, the user profile (user ID, age, city, etc.), the item profile (category, item ID, etc.), and the context information (position, weekday, etc.) \cite{zhu2020fuxictr}.
The input data for deep recommender systems is commonly in tabular format, i.e., numerical, categorical, or multi-valued features of multiple fields, as opposed to other forms of information like texts or images. The sample size of the tabular data is typically immense, with a highly sparse feature space \cite{cheng2016wide}. It is common to apply one-hot or multi-hot encoding to map the raw features as binary features into high-dimensional feature space. For categorical feature field $i$, binary feature $\mathbf{x}_i$ is obtained by one-hot encoding. For numeric feature field, the numeric values are bucketed into discrete features manually (e.g. [0, 0, 0, 1] for $age \in [0, 14]$) or by training decision trees (e.g., GBDT \cite{he2014practical}), and encoded as categorical feature field. Multi-valued fields are encoded by multi-hot encoding. The concatenation of binary features consists of a user-item interaction data instance $\mathbf{x} = [\mathbf{x}_1, \mathbf{x}_2, \cdot \cdot \cdot, \mathbf{x}_f]$ \cite{zhao2021autoloss}:
\begin{equation*}
\underbrace{[1, 0,\cdot\cdot\cdot, 1]}_\text{$\mathbf{x}_{1}$: user ID} \underbrace{[0, 1,\cdot\cdot\cdot, 0]}_\text{$\mathbf{x}_{2}$: age} \underbrace{\cdot\cdot\cdot\cdot\cdot \ \cdot}_\text{other fields}\underbrace{[1, 1, 1,\cdot\cdot\cdot, 0]}_\text{$\mathbf{x}_{f}$: item ID}
\end{equation*}
where $f$ is the number of feature fields and $ \mathbf{x}_i$ is the binary vector for $i^{th}$ feature field.

\subsubsection{Embedding Layer} 
The binary vectors are usually high-dimensional and sparse, which can be transformed into low-dimensional and dense vectors by feature embeddings. The embedding process is presented as follows:  

% Typically, the original input features are categorical, numerical or multi-valued. One-hot and multi-hot encoding are introduced to transform the raw input features into binary vectors. As for numerical data, they are divided into several buckets, and every bucket is assigned with a binary vector. All the binary vectors concatenate the user-item interaction data $\mathbf{x} = [\mathbf{x}_1, \mathbf{x}_2, \cdot \cdot \cdot, \mathbf{x}_f]$, where $f$ is the number of feature fields and $ \mathbf{x}_i$ is the binary vector for $i^{th}$ feature field. The binary vector $\mathbf{x}_i$ is usually high-dimensional and sparse, 

For the binary feature $\mathbf{x}_i$ generated from categorical or numeric feature field, feature embedding $\mathbf{e}_i$ is obtained by:

\begin{equation}
\mathbf{e}_{i} = \mathbf{E}_i \mathbf{x}_i
\label{equa:embedding1}
\end{equation}
where $\mathbf{E}_i \in \mathbb{R}^{d \times n_i}$ is the embedding matrix transforming the $i^{th}$ binary feature to condensed feature embedding $\mathbf{e}_i$. For $i^{th}$ feature field, $d$ is the size of pre-defined low-dimensional embedding and $n_i$ is the number of distinctive feature values. 

For multi-valued feature field $h$, $\mathbf{E}_h$ is the embedding matrix. The embedding is obtained by:
\begin{equation}
\mathbf{e}_{h} = \mathbf{E}_h [\mathbf{x}_{h_1}, \mathbf{x}_{h_2},\cdot\cdot\cdot, \mathbf{x}_{h_{u_h}}]
\label{equa:embedding2}
\end{equation}
where every feature is represented as a sequence, $\mathbf{x}_{h_{u_h}}$ as the one-hot encoded binary vector, and $u_h$ is the maximal length of the sequence. The embedding $\mathbf{e}_h \in \mathbb{R}^{ d \times u_h}$ can be aggregated to a $d$ dimensional vector by mean or sum pooling.

The output of embedding layer is the concatenation of all feature embeddings:
\begin{equation}
\mathbf{E} = [\mathbf{e}_1, \mathbf{e}_2,\cdot\cdot\cdot,\mathbf{e}_f]
\end{equation}

% The feature embedding matrix $\mathbf{E} = \{\mathbf{e}_1, \mathbf{e}_2,...,\mathbf{e}_f\}$ 

% DEFINITION 1. (Feature)

% DEFINITION 1. (Feature Field)

% DEFINITION 1. (Feature Embedding)

% DEFINITION 1. (Feature Component) $\mathbf{e}_i = [e^1_i,e^2_i,...,e^{s_i}_i]$, where $s_i$ is the embedding size for $\mathbf{e}_i$

\subsubsection{Interaction Layer}
After the raw features are mapped in the low-dimensional space, an interaction layer is proposed to capture the feature interaction information between different feature fields. There are two types of feature interactions: explicit feature interactions and implicit feature interactions. Explicit feature interactions implement interaction functions among specific features, which are interpretable, and people acknowledge which features play an essential role in the model performance. Implicit feature interactions use multi-layer perceptron (MLP) to learn the non-linear information from all the feature embeddings.

Based on the convention definition \cite{lian2018xdeepfm}, and the mathematical definition proposed in \cite{lyu2021memorize}, the result of explicit $p^{th}$ order $(1\leq p \leq f)$ of feature interaction is obtained by the feature embedding group $\mathcal{P} = \{\mathbf{e}_i\}_{i=c_1,c_2,\cdot\cdot\cdot,c_p}$:
\begin{equation}
\mathbf{e}_{\mathcal{H}} = o^{(p-1)}(\cdot\cdot\cdot(o^{(1)}(\mathbf{e}_{c_1},\mathbf{e}_{c_2})),\cdot\cdot\cdot,\mathbf{e}_{c_h})
\end{equation}
where every $e_i$ in $\mathcal{P}$ is searched from the concatenation of all feature embeddings $E$, and $o^{(p-1)}(\cdot)$ is a feature interaction function, commonly designed by human experts. For example, Factorization Machines (FM) \cite{rendle2010factorization,9101475} implements the inner product of feature embeddings to explicitly model the $2^{nd}$ order feature interactions, and define $1^{st}$ order interaction as the binary vector $\mathbf{x}_i$. In this scenario, the output of the interaction layer $\boldsymbol{l}_{FI}$ will be the output of FM:

\begin{equation}
output_{FM} = <\boldsymbol{w}, \mathbf{x}> + \sum^{m}_{i=1}\sum^{m}_{j>i} <\mathbf{e}_i,\mathbf{e}_j>
\end{equation}
where $\boldsymbol{w}$ is the weight for binary vector $\mathbf{x}$, $\mathbf{e}_i$ is the low dimensional feature embedding of $i^{th}$ field, and $<\mathbf{e}_i,\mathbf{e}_j>$ is the inner product of vector $\mathbf{e}_i$ and vector $\mathbf{e}_j$. In theory, FM can explicitly model any order of feature interactions by the inner product of the corresponding feature embeddings. However, high order ($p^{th} $ order with $p \geq 3$) feature interaction introduces exponentially grown computation with respect to $p$. 

% \textbf{Factorization models}
% \begin{equation}
% Output_{fm} = \sum^{m}_{i=1} w_i \mathbf{e}_i + \sum^{m}_{i=1}\sum^{m}_{j>i}w_{ij} <\mathbf{e}_i,\mathbf{e}_j>
% \end{equation}

Multi-layer perceptron (MLP) can both learn the implicit feature interactions and integrate different orders of feature interaction and various types of embeddings, by extracting non-linear information with fully-connected layers and activation functions. The output of every layer $\boldsymbol{h}_{l+1}$ is:

\begin{equation}
\boldsymbol{h}_{l+1} = \sigma(\boldsymbol{W}_l \boldsymbol{h}_l + \boldsymbol{b}_l)
\label{equ:MLP}
\end{equation}
where $\sigma(\cdot)$ is the activation function, $\boldsymbol{W}_l$ is the weight, $\boldsymbol{h}_l$ is the outputs of the previous layers, and $\boldsymbol{b}_l$ is the bias. In many hand-crafted models, the output of MLP is combined with other embeddings before being taken as the output of the multi-interaction ensemble layer. DeepFM \cite{guo2017deepfm} concatenates the output of $2^{nd}$ order feature interaction $\boldsymbol{l}_{FM}$ and the output of MLP on the embedding matrix, denoted as $\boldsymbol{l}_{FI} = concat(\boldsymbol{l}_{FM}, MLP(\mathbf{e}))$. IPNN \cite{qu2016product} feeds explicit feature interactions and embedding matrix to MLP, and the output of feature interaction layer is denoted as $\boldsymbol{l}_{FI} = MLP(\boldsymbol{l}_{FM}, \mathbf{e}))$. 

\subsubsection{Prediction Layer}
The prediction layer yields the prediction $\hat{y}$ based on the output of the feature interaction layer:
\begin{equation}
\hat{y} = \sigma(\boldsymbol{W}_o \boldsymbol{l}_{ensemble} + \boldsymbol{b}_o)
\end{equation}
where  $\boldsymbol{W}_l$ is the weight, and $\boldsymbol{b}_l$ is the bias. The specific recommendation task drives the choice of activation function $\sigma(\cdot)$. For instance, $sigmoid$ is preferred for the binary classification task \cite{guo2017deepfm}, whereas $softmax$ is selected for multi-class classification \cite{song2019autoint}. The loss is calculated for the backpropagation steps based on the ground truth label $y$:
\begin{equation}
\mathcal{L} = \ell(y,\hat{y})
\end{equation}
where $\ell$ is the loss function, such as cross-entropy and mean-squared-error, usually determined by human experts \cite{zhao2021autoloss}.

\subsection{Neural Architecture Search (NAS)}
Neural Architecture Search (NAS) \cite{elsken2019neural,pmlr-v80-pham18a} is proposed to search for the data-oriented and task-specific ideal deep learning architecture from the search space, alleviating considerable human effort in the architecture design procedure. Most works explore well-performed convolutional neural network architectures for different tasks like image classification \cite{chen2018searching} and Natural language processing \cite{klyuchnikov2020bench}. The NAS methods are determined by two significant factors: search space and search strategy. The set of all possible architectures is search space, which is broad enough to encompass existing well-performed architectures but still maintains a reasonable size to prevent increasing search costs. Within the search space, the search strategy efficiently searches for the preferred architecture and is expected to balance exploration and exploitation.

There are primarily three kinds of methodologies in the NAS field. (1) Sample-based approaches \cite{cai2018efficient} explore new architecture by selecting from search space or mutating existing promising ones. (2) Reinforcement learning-based approaches \cite{liu2017hierarchical,real2017large} implement a recurrent neural network as a policy controller to produce a sequence of actions to determine the architecture design. (3) Gradient-based approaches \cite{xie2018snas,liu2018darts} convert the discrete search space to continuous and optimize the search architecture with the gradient descent calculated from the performance on the validation set. Differentiable Architecture Search (DARTS) employs continuous relaxation to search over the non-differentiable and discrete search space as a favored gradient descent-based NAS method. 

\subsubsection{Continuous Relaxation} Let $\mathcal{\hat{O}} = \{\hat{o}_{j}(\cdot)\}$ represents the set of operations (e.g., the set of feature interaction candidate functions). To convert the search space continuous, a weight vector $\boldsymbol{\alpha}^{(i)} = [\alpha^{(i)}_1,\cdot\cdot\cdot,\alpha^{(i)}_{|\mathcal{\hat{O}}|}]$ indicates the contribution of individual operation to the model performance, where $1\leq i \leq I$, and $I$ is the number of positions waiting to be put a selected operation. The original one operation $\hat{o}^{(i)}(\cdot)$ at position $i$ is replaced by the mixed operation $\bar{o}^{(i)}(\cdot)$  with a softmax over all candidates \cite{maddison2016concrete}:
\begin{equation}
\bar{o}^{(i)}(\cdot) = \sum_{j=1}^{|\mathcal{\hat{O}}|}\frac{\exp(\alpha_j^{(i)})}{\sum_{n=1}^{|\mathcal{\hat{O}}|}\exp(\alpha_n^{(i)})}\hat{o}_{j}(\cdot)
\end{equation}

The task of searching for a well-performed discrete architecture with needed operations at $I$ positions is altered to jointly learn the architecture set $\mathcal{A}=\{\boldsymbol{\alpha}^{(i)}\}_{1 \leq i \leq I}$, and the weight set $\mathcal{W}$ of all operators. After learning procedure, the operator at $i^{th}$ position of the outputted discrete architecture is the one with the largest weight in the vector $\boldsymbol{\alpha}^{(i)}$:
\begin{equation}
    \hat{o}^{(i)}(\cdot) = \arg \max_{1 \leq j \leq |\mathcal{\hat{O}}|} \alpha_j^{(i)}
\end{equation}

\subsubsection{Gumbel-Softmax Operation} Like the above continuous relaxation, the Gumbel-Softmax operation substitutes a differentiable sample for the original non-differentiable categorical variable with a Gumbel-Softmax distribution \cite{jang2016categorical}. Thus stochastic neural networks can perform backpropagation through examples. Given the continuous distribution $\boldsymbol{\alpha}=[\alpha_1,\cdot\cdot\cdot,\alpha_n]$ over the candidates, a hard selection $z$ is drawn by the Gumbel-Max trick \cite{gumbel1954statistical}:

\begin{equation}
    z =  \text{ one\_hot } \left(\arg \max_{1 \leq i \leq n} \left(\log \alpha_i + g_i\right)\right)
\label{equa:Gumbel-Softmax}
\end{equation}
where $\{g_i\}_{1 \leq i \leq n}$ are independent and identically distributed (i.i.d) noise samples drawn from $-\log(-\log(u_i))$ and $u_i \sim Uniform(0,1)$.
However, the sampling is non-differentiable due to $\arg \max$ operation. The softmax function replaces $\arg \max$ as a continuous and differentiable approximation. Gumbel-Softmax generates $p_i$, the probability of selecting the $i^{th}$ candidate as:
\begin{equation}
    p_i = \frac{\exp\left( \frac{\left(\log\left(\alpha_i\right)+g_i\right)}{\tau}\right)}{\sum^n_{j=1} \exp\left( \frac{\left(\log\left(\alpha_i\right)+g_j\right)}{\tau}\right)}
\end{equation}
where $\tau$ is the temperature parameter that controls the smoothness of the operation. The Gumbel-Softmax operation's output turns into a one-hot vector as $tau$ gets closer to zero.

\subsubsection{Bi-level Optimization} Similar to sample-based NAS \cite{cai2018efficient} and reinforcement learning-based NAS \cite{liu2017hierarchical,real2017large} using the model performance over the validation set as the fitness or reward, DARTS \cite{liu2018darts} utilizes the validation set to guide the learning procedure of $\mathcal{A}$ and $\mathcal{W}$ by optimizing the validation loss and training loss in a gradient descent manner. The output of NAS $\mathcal{A}^{*}$ is acquired by minimizing the loss on the validation set, while the weight set $\mathcal{W}^*$ is attained by minimizing the loss on the training set, which can be formulated as a bi-level optimization problem \cite{anandalingam1992hierarchical,colson2007overview}:

% \begin{equation}
% \begin{split}
% & \min_{\mathcal{A}} \mathcal{L}_{val}(\mathcal{W}^*(\mathcal{A}),\mathcal{A}) \\
% & \text{s.t.} \ \mathcal{W}^*(\mathcal{A}) = \arg \min_{\mathcal{W}}\mathcal{L}_{train}(\mathcal{W},\mathcal{A})
% \end{split}
% \end{equation}

\begin{subequations}
\begin{align}
& \min_{\mathcal{A}} \mathcal{L}_{val}(\mathcal{W}^*(\mathcal{A}),\mathcal{A}) \\
\label{equa:bi1}
& \text{s.t.} \ \mathcal{W}^*(\mathcal{A}) = \arg \min_{\mathcal{W}}\mathcal{L}_{train}(\mathcal{W},\mathcal{A})
\end{align}
\label{equa:bi2}
\end{subequations}

where $\mathcal{L}_{train}$ and $\mathcal{L}_{val}$ represent the validation loss and training loss. $\mathcal{A}$ is the upper-level parameter and $\mathcal{W}$ is the lower-level parameter.

\subsubsection{Architecture Gradient Approximation} After constructing the NAS problem as a bi-level optimization, DARTS introduces a straightforward approximation strategy to overcome the costly internal variable optimization in equation \ref{equa:bi2}. Instead of solving the internal optimization entirely by training to convergence, $\mathcal{W}(\mathcal{A})$ is approximated by varying $\mathcal{W}$ with one training step only \cite{liu2018darts}. Thus this trick can also be called a one-step approximation:
\begin{subequations}
\begin{align}
    & \nabla_{\mathcal{A}}\mathcal{L}_{val}(\mathcal{W}^*(\mathcal{A}),\mathcal{A}) \\
    \approx &\nabla_{\mathcal{A}}\mathcal{L}_{val}(\mathcal{W} - \xi \nabla_{\mathcal{W}}\mathcal{L}_{train}(\mathcal{W},\mathcal{A}), \mathcal{A})
\end{align}
\label{equa:gra}
\end{subequations}
where $\xi$ represents the learning rate for one step in the internal variable optimization, and $\mathcal{W}$ indicates the contemporary weights acquired by the optimization. If $\mathcal{W}$ reaches the local optimum for the internal optimization, $\nabla_{\mathcal{W}}\mathcal{L}_{train}(\mathcal{W},\mathcal{A}) = 0$ and equation \ref{equa:gra} degenerates to:
\begin{equation}
\nabla_{\mathcal{A}}\mathcal{L}_{val}(\mathcal{W}^*(\mathcal{A}),\mathcal{A}) \approx \nabla_{\mathcal{A}}\mathcal{L}_{val}(\mathcal{W},\mathcal{A})
\end{equation}

After the approximate architecture gradient is applied with the chain rule, it becomes:
\begin{equation}
\nabla_{\mathcal{A}}\mathcal{L}_{val}(\mathcal{W}^{\prime}),\mathcal{A}) - \xi\nabla_{\mathcal{A},\mathcal{W}}^{2}\mathcal{L}_{train}(\mathcal{W},\mathcal{A})\nabla_{\mathcal{W}^{\prime}}\mathcal{L}_{val}(\mathcal{W^{\prime}}, \mathcal{A})
\label{equ:first-orderapp}
\end{equation}
where $\mathcal{W}^{\prime}=\mathcal{W}-\xi \nabla_{W}\mathcal{L}_{train}(\mathcal{W},\mathcal{A})$ indicates the weights of the one-step forward model. $\xi = 0$ accelerates the optimization procedure by neglecting the second order derivative, which is called first order approximation. Second order approximation refers to the scenario where $\xi > 0$. The choice of different approximation methods is the trade-off between accuracy and efficiency.

% Table generated by Excel2LaTeX from sheet 'Sheet1'
\begin{table}[htbp]
  \centering
  \caption{Frequently used notations.}
    \begin{tabular}{ll}
    \toprule
    \textbf{Notations} & \textbf{Descriptions} \\
    \midrule
    \midrule
    $\mathcal{U}/\mathcal{I}$ & set of users/items \\
    $\mathcal{X} $ & set of user-item examples \\
    $\mathcal{O} $ & set of feature interaction functions \\
    $|\mathcal{U}|$ & number of users \\
    
    $\mathbf{f}_i$ & feature field $i$ \\
    $ \mathbf{x}_i$ & binary vector for $i^{th} $ feature field \\
    $\mathbf{x} $ & one user-item example \\
    $ \mathbf{e}_i$ & feature embedding for $i^{th} $ binary vector \\
    $\mathbf{d}$ & vector of mixed embedding dimensions \\
    $\mathbf{E}$   & embedding matrix  \\

    $F$   & number of features fields \\
    $d$ & pre-defined embedding dimension \\
    $d_i$ & embedding dimension for $i^{th}$ feature field \\
    % $\hat{d}$ & base embedding dimension in embedding alignment \\
    
    % $\mathbb{R}$ & real number \\
    $n_i$ & number of distinctive feature values for feature field $i$\\
    
    $K$   & pre-defined highest order for interaction \\
    
    $A \otimes  B$       &   Interaction between feature A and feature B  \\
    $\mathbf{e}_i \circ \mathbf{e}_j $ & element-wise product between $\mathbf{e}_i$ and $\mathbf{e}_j$ \\
    $w$   & trainable weight \\
    $o(\cdot)$ & feature interaction function \\
    $\sigma(\cdot)$ & activation function \\
    $<\mathbf{e}_i,\mathbf{e}_j>$ & inner product of $\mathbf{e}_i$ and $\mathbf{e}_j$ \\
    $\ell(\cdot)$ & loss function \\
    $\mathcal{L}$ & calculated loss value \\
    \bottomrule
    \end{tabular}%
  \label{table:sign}%
\end{table}%

\section{Automated Feature Selection Search (Auto-FSS)}
\label{sec:autofss}
As mentioned above, the input for deep recommender systems is binary vectors of feature fields. Most existing works \cite{guo2017deepfm,he2017neural} collect and use as many features as possible,  regardless of whether the features are helpful for recommendation or not. This paradigm frequently calls for extra computations cost for feature embedding learning, additional inference time, and sub-optimal performance caused by the redundant or irrelevant features \cite{nadler2005prediction}. Therefore, selecting a subset of principal feature fields for deep recommender systems is highly demanded. The traditional feature selection methods such as hand-crafted by human experts, grid search \cite{glorot2010understanding,hinton2012practical}, or filter methods \cite{yu2003feature} cannot be seamlessly integrated with deep recommender systems. For instance, filter methods omit the connection between subsequent models and feature selection. Therefore, automatically selecting features specifically as the input of recommendation models has attracted much attention in recent years, which is termed the automated feature selection search (Auto-FSS) in this paper. Auto-FSS faces the following two challenges. (1) \textbf{Huge search space:} Real-world recommender systems possess a vast number of unique features (e.g., more than a billion user IDs on YouTube) \cite{covington2016deep}.  These features and their cross features that are generated by operations over the unique features define the huge search space. Efficiently performing a search on the huge search space is a challenging problem. (2) \textbf{Dynamic feature significance:} In the recommendation tasks, the significance of a particular feature field may vary widely for different user-item interaction instances. Discovering the same useful feature subset for all instances limits the recommendation performance. It is worth mentioning that we only survey Auto-FSS for deep recommender systems as shown in Table~\ref{table:autofss}, focusing on the above challenges since our main scope is AutoRecSys.

\begin{table*}[htbp]
  \centering
  \caption{Summary of Auto-FSS methods.}
  \begin{adjustbox}{max width=\textwidth}

\begin{tabular}{ccccc}
\toprule
\textbf{Methods} & \textbf{Search Space} & {\textbf{Search Strategies}} & {\textbf{Search Level}} & {\textbf{Tasks}} \\
\midrule
\midrule
{AutoField (WWW'2022) \cite{wang2022autofield}} & {$2F$} & {Gradient-based} & {Raw Feature} & {CTR} \\
\midrule
{AdaFS (KDD'2022) \cite{lin2022adafs}} & {$F$} & {MLP} & {Raw Feature} & {CTR} \\
\midrule
{GLIDER(ICLR'2020)  \cite{tsang2019feature}} & {$2^{2^F}$} & {Gradient-based} & {Generated Feature} & {CTR} \\
\midrule
{AutoCross (KDD'2019) \cite{luo2019autocross}} & {$(\frac{F^2}{2})^n$} & {Beam Search} & {Generated Feature} & {CTR} \\
\bottomrule
\end{tabular}%

    \end{adjustbox}
  \label{table:autofss}%
\end{table*}%

\textbf{AutoField} \cite{wang2022autofield} combines the feature selection process with the downstream recommendation tasks by implementing $F$ two-dimensional controller vectors $[\alpha^{1}_{f}, \alpha^{0}_f]$ to learn the contribution of each feature to the CTR prediction, where $\alpha^{0}_{f}$ denotes the possibility of neglecting a feature filed, and $\alpha^{1}_{f}$ represents the possibility of choosing a feature filed, $\alpha^{1}_{f}+\alpha^{0}_f = 1$, and $1 \leq f \leq F$. After parameters of controller vectors and deep recommender models are jointly learned by DARTS in a bi-level optimization manner, the Gumbel-Max trick simulates the hard feature selection process according to the controller parameters.

Following the idea of the controller in AutoField, \textbf{AdaFS} \cite{lin2022adafs} provides an adaptive feature selection method for dynamic data instances rather than a static set. It utilizes a controller network with several fully-connected layers to output weights $\{\alpha^{m}_{f}\}_{1\leq f \leq F}$, revealing the importance of different feature fields for a data instance. Before the feature embedding is fed to the MLP as the input, BatchNorm \cite{ioffe2015batch} is implemented to make the feature embedding $\mathbf{e}_f$ with various magnitude comparable: 
\begin{equation}
\hat{\mathbf{e}}^m_f = \frac{\mathbf{e}^m_f - Mean^f_{\mathcal{B}}}{\sqrt{Var^{f}_{\mathcal{B}}+ \epsilon}}
\end{equation}
where $m$ represents a index for $m^{th}$ data instance in the batch, $Mean^f_{\mathcal{B}}$ is a mini-batch mean, $Var^{f}_{\mathcal{B}}$ represents a variance, and $\epsilon$ is a small constant.

In addition to selecting relevant features from raw feature sets, some literature finds and produces useful combinatorial features (i.e., cross features), such as statistical features. \textbf{GLIDER} \cite{tsang2019feature} utilizes gradient-based Neural Interaction Detection (NID) \cite{tsang2018detecting} to detect statistical feature interactions that span globally across multiple data instances from a source recommender model with a perturbation model called LIME \cite{ribeiro2016should}. Then, GLIDER explicitly encodes the generated features (i.e., searched global interaction) into a target recommendation model. The target recommendation model can be any classical recommender system, and the generated features are explored in $2^{2^{f}}$ search space.

\textbf{AutoCross} \cite{luo2019autocross} enables feature selection from generated cross features, and performs the search in a tree-structured search space by implementing greedy beam search. It is different from feature interactions, as the output of AutoCorss is the useful feature sets, which can be fed to different recommendation models such as  Wide\&Deep \cite{cheng2016wide}. The search space is tree-structured with the original feature set $\mathcal{F}$ as the root and other nodes as the feature interaction set. If one node contains $f^{\prime}$ feature interactions, including original features, the number of children of that node is $\frac{f^{\prime}(f^{\prime}-1)}{2}$. Every child node is the parent node set, adding one feature interaction from the parent node set. The size of the search space is $O((\frac{F^2}{2})^n)$, growing exponentially with the maximum number of generated feature interactions $n$.

To deal with the immense search space, the beam search, as a greedy strategy, is implemented in the tree-structured search space to traverse it from the root efficiently, by only exploring the most promising child node after evaluating all the children feature interaction sets for one node. In this greedy manner, it expands linearly with parameter $n$, since only $nF^2$ nodes will be evaluated in the $(\frac{F^2}{2})^n$ search space. 

In the feature set evaluation stage, field-wise logistic regression is applied to accelerate the process. It approximates the actual performance with mini-batch gradient descent. The model prediction is:
\begin{equation}
P(y=1|\mathbf{x})= Sigmoid(\boldsymbol{w}_{new}\mathbf{x}_{new} + \boldsymbol{w}_c\mathbf{x}_c)
\end{equation}

where $\mathbf{x}_{new} $ and $\mathbf{x}_c $ represent newly added interactive feature and features in the current set, respectively. Only the weight of the newly added feature interaction $\boldsymbol{w}_{new}$ will be learned in the model training stage, while weights of current feature interactions $\boldsymbol{w}_{c}$ stay fixed. 

In the data processing stage, AutoCross proposes multi-granularity discretization, which discretizes one numerical feature with several levels of granularity rather than pre-defined granularity. It evaluates different discretized features, and only the best one is kept.

We analytically compare the above Auto-FSS methods as follows. AutoField and AdaFs search in the raw feature level within compact search spaces, while others search in the high-order generated cross feature level. There is no absolute winner among these two categories. Searching for raw features enables the task-specific subsequent recommender systems to discover the interactions and correlations within the original input features. It works well when the recommendation systems are expressive and have adequate computation resources. When the data distribution is rapidly changing in the scenario, or the significance of the high-order cross feature alters, the former is preferred since the downstream recommendation system can be finetuned, and the re-search procedure for the latter is time-consuming. 

Within the raw feature search category, AdaFs selects different features for different data instances to address the second challenge, while AutoField outputs constant feature sets. Within the generated feature search category, AutoCross searches the explicit high-order feature interactions and can be fully deployed on the distributed systems to fit industrial needs. Both traditional recommender systems and deep neural models can be implemented after AutoCross constructs the feature interaction set. The drawback is also apparent. Exploring the high-order interaction feature space in a trial-and-error manner to prune the search pace leads to sub-optimal results.

% For the future direction, we would like to see more Auto-FSS methods for sequential recommendation systems, which play an essential role in industrial recommendation circumstances. New contents and new labels are created from individuals or companies daily \cite{chen2018sequential}. It means Auto-FSS should consider not only the data distributions but also the users' interest shifts and newly emerged features. How to quickly evaluate the new incoming features based on the existing feature selection model still needs more discussions and experiments from the community.

\section{Automated Embedding Dimension Search (Auto-EDS)}
\label{sec:autoeds}
\begin{figure*} 
\centering
\includegraphics[width=\textwidth]{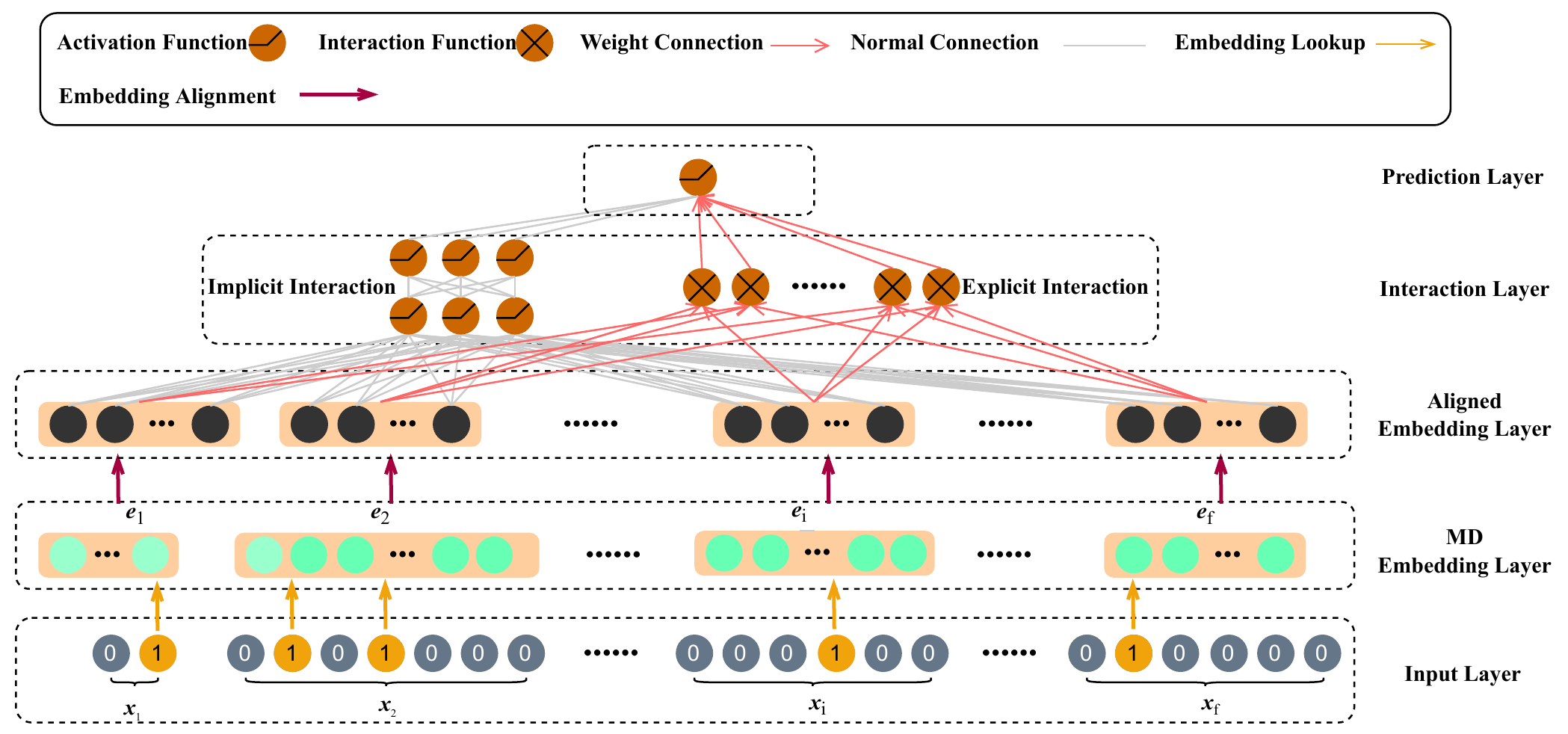}
\caption{The framework of Auto-EDS.}
\label{fig:autoeds}
\end{figure*}
As mentioned above, the typical inputs of recommender systems involve many feature fields, and each field consists of a certain number of unique features ranging from a few to hundreds of millions.
Since these original features are generally encoded as high-dimensional and sparse vectors, most modern deep recommender systems \cite{cheng2016wide, covington2016deep,rendle2010factorization, guo2017deepfm} map them into the low-dimensional and dense feature representations (a.k.a., embeddings) for better capturing the implicit feature information. However, most of these methods assign a uniform embedding dimension for all features, suffering from the following issues. (1) \textbf{Resources consumption:} The huge number of parameters in the embedding matrix consume a large amount of storage and computational resources of the model. (2) \textbf{Unappealing performance:} The features generally follow a long-tail distribution in recommender systems \cite{park2008long}, so setting the same dimension for head features and tail features may lead to sub-optimal performance. In particular, the high-frequency features need more parameters to express their rich information, and over-parameterizing the low-frequency features could cause overfitting due to the limited training data.

Thus, assigning embedding dimensions to each feature (field) automatically has attracted much attention in recent years, which is termed the automated embedding dimension search (Auto-EDS) in this paper. Concretely, as shown in Fig. \ref{fig:autoeds}, we introduce an overview of Auto-EDS architecture. The key component of Auto-EDS is the mixed dimension (MD) embedding layer, which consists of variable embedding sizes for each feature (fields). Then these MD embeddings are aligned into the same dimension via alignment operators (e.g., projection matrices \cite{ginart2021mixed}) in the aligned embedding layer in order to satisfy the further operations (e.g., the dot product) in the interaction layer.

Existing methods in this research line could be categorized into heuristic, hyper-parameter optimization (HPO), and embedding pruning methods. We summarize these methods in Table~\ref{table:autoeds}.
% Table generated by Excel2LaTeX from sheet 'Sheet1'

\subsection{Heuristic Methods}
The heuristic methods generally assign embedding dimensions for each
feature (field) based on the pre-defined human rules. For example, \textbf{MDE} \cite{ginart2021mixed} introduces a mixed dimension (MD) embedding layer which assigns the embedding dimension based on the feature's popularity. In particular, the MD layer consists of $F$ blocks corresponding to $F$ feature fields for the CTR task, and each block is defined by the embedding matrix $\mathbf{E}_{i} \in \mathbb{R}^{n_{i} \times d_{i}}$ and the projection matrix $\mathbf{P}_{i} \in \mathbb{R}^{d_{i} \times \hat{d}}$, respectively. The former stores embedding vectors for $i^{th}$ block (field) with dimension $d_{i}$, and $n_{i}$ is the number of unique features in the field. The latter $\boldsymbol{P}_{i}$ aligns the dimension into a base dimension $\hat{d} \ge d_{i}$ for the further feature operations (e.g., the inner product) requiring consistent embedding dimensions. Thus, the question is how to assign mixed embedding dimensions $\mathbf{d} = [d_{1},...,d_{F}]$ for $F$ blocks. To this end, MDE first defines the block-level probability vector $\mathbf{p} = [\frac{1}{n_{1}},...,\frac{1}{n_{F}}]$, and then the mixed embedding dimensions $\mathbf{d}$ could be obtained as follows:
\begin{equation}
    \mathbf{d} = \hat{d}||\mathbf{p}||_{\infty}^{-\alpha}\mathbf{p}^{\alpha}
\end{equation}
where $\alpha$ is the hyperparameter controlling the degree of popularity influencing the embedding dimension. 

Although such heuristic methods provided a simple scheme for assigning mixed embedding dimensions, the assumptions, i.e., the spectral decay following the power law, behind it were not guaranteed always to be satisfied, thus limiting its generalization on complex tasks.

\subsection{HPO Methods}
Inspired by the recent success of neural architecture search (NAS) \cite{zoph2016neural,elsken2019neural} for automatically searching architectures for deep neural networks, another research line in Auto-EDS is to consider it as the hyper-parameter optimization (HPO) problem that searches embedding dimensions from a pre-defined candidate dimension set.

For example, \textbf{NIS} \cite{joglekar2020neural} (Neural Input Search) is the first HPO-based Auto-EDS work to automatically learn embedding dimensions and vocabulary sizes for each unique feature. In particular, it first sorts the $v$ unique features in decreasing order based on their frequency in the dataset resulting in a $\mathbf{E} \in \mathbb{R}^{ v \times \hat{d}}$ embedding matrix, where $\hat{d}$ is the pre-defined maximum embedding dimension. Then, the $\mathbf{E}$ is divided into a grid of $S \times T$ submatrices (embedding blocks) with $S > 1$ and $T > 1$, and the $(s,t)^{th}$ submatrix is of size $\hat{v}_{s} \times \hat{d}_{t}$ such that $v= \sum_{s=1}^{S}\hat{v}_{s}$ and $d = \sum_{t=1}^{T}\hat{d}_{t}$. In order to learn a MD embeddings, inspired by ENAS \cite{pmlr-v80-pham18a}, a controller is employed to make a sequence of $T$ choices, and each choice is an $\hat{s}_{t} = \{1,...,S\} \cup \{0\} $, where $\hat{s}_{t}=0$ means that the $\hat{d}_{t}$-dimensional embedding is removed. In this way, the search space size is $(S+1)^{T}$, and the reward of the controller is defined by the combination of the quality reward (i.e., the metric values of the model evaluation on the validation set.) and the memory cost. Finally, the model is first trained by a warm-up phase to ensure that embedding blocks are expressive; after that, the main model (i.e., the recommendation model) and the controller are trained alternatingly by A3C \cite{mnih2016asynchronous}.

Inspired by the differentiable NAS (DARTS) \cite{liu2018darts}, \textbf{DNIS} \cite{cheng2020differentiable} proposes a differential NIS framework to improve the search efficiency. Specifically, unlike NIS, which searches for embedding dimensions from pre-defined discrete dimension sets, DNIS argues that this restriction will hurt the flexibility of dimension selection and relax the search space to be continuous via the soft selection layer. In particular, for $v$ unique features, it uses $v$ binary dimension index vectors $\mathbf{\hat{d}}$ to maintain the ordered locations of the feature's existing dimensions from the pre-defined dimension set $\{1,\cdots,d\}$, and then uses the similar feature sorting method as NIS to divide $v$ features into $S$ blocks such that features within the same block share the same binary dimension index vector. Thus, the total search space is $2^{Sd}$. To learn $\mathbf{\hat{d}}$ efficiently, the $\mathbf{\hat{d}}$ is relaxed to be a continuous variable $\mathbf{\bar{d}}$ within the range of $[0,1]$, named the soft selection layer, which is inserted between the embedding layer and the interaction layer. In this way, the relaxed dimension vectors could be jointly optimized with the embedding matrix by gradient descent. Furthermore, the gradient normalization operation is utilized to avoid numerical overflow. After training, the final output embedding matrix $\hat{E}$ is obtained as follows:
\begin{equation}
    \mathbf{\hat{E}} = \mathbf{E} \circ \mathbf{\bar{d}}
\end{equation}
where $\circ$ is the element-wise product. Finally, to obtain the MD embeddings, we could prune the $\mathbf{\hat{E}}$ by a pre-defined threshold.

\textbf{AutoEmb} \cite{zhao2020autoemb} proposes an AutoML-based end-to-end framework in streaming recommendations that could automatically and dynamically select embedding dimensions for users/items based on their popularity changes. Specifically, each user and item are defined in the $w$ embedding spaces with embeddings $\{\mathbf{e}^{1} \in \mathbb{R}^{d_{1}},\cdots,\mathbf{e}^{w} \in \mathbb{R}^{d_{w}}\}$ ,$\{\mathbf{h}^{1} \in \mathcal{R}^{  d_{1}},\cdots,\mathbf{h}^{w} \in \mathbb{R}^{  d_{w}} \}$, respectively, where $d_{1}<\cdots<d_{w}$ are the corresponding embedding dimensions. Then, embeddings in different spaces are unified into the same space via the linear transform as follows:
\begin{equation}
    \hat{\mathbf{e}}_{i} = \mathbf{W}_{i} \mathbf{e}_{i} + \mathbf{b}_{i}
\end{equation}
where $\mathbf{W}_{i} \in \mathbb{R}^{d_{w} \times d_{i}}$ and $b_{i} \in \mathbb{R}^{  d_{w}}$ are the weight matrix and bias vector, respectively. In order to make the magnitude of the transformed embeddings comparable, the Batch-Norm \cite{ioffe2015batch} with the Tanh activation function \cite{karlik2011performance} is utilised to normalize the transform embeddings resulting in the magnitude-comparable user embedding vectors $\{\bar{\mathbf{e}}_{1} \in \mathbb{R}^{d_{w}},\cdots,\bar{\mathbf{e}}_{w} \in \mathbb{R}^{d_{w}}\}$ and item embedding vectors $\{\bar{\mathbf{h}}_{1} \in \mathbb{R}^{d_{w}},...,\bar{\mathbf{h}}_{w} \in \mathbb{R}^{d_{w}}\}$. Thus, the search space is $2^{(U+I)w}$, where $U$ and $I$ are the numbers of users and items, respectively. Furthermore, two MLP-based controller networks are used to choose dimensions from the above candidates for users and items based on their popularity and contextual information separately. In order to make the whole framework end-to-end differentiable, the soft selection layer that employs weighted sum of $\{\bar{\mathbf{e}}_{1} \in \mathbb{R}^{d_{w}},\cdots,\bar{\mathbf{e}}_{w} \in \mathbb{R}^{d_{w}}\}$ is utilised to obtain the final representations $\bar{\mathbf{e}^{*}}$ and $\bar{\mathbf{h}^{*}}$ of users and items as follow:  
\begin{equation}
\begin{aligned}
&\bar{\mathbf{e}^{*}}=\frac{1}{w} \sum_{i=1}^{w} \alpha_{i} \cdot \bar{\mathbf{e}}^{i} \\
&\bar{\mathbf{h}^{*}}=\frac{1}{w} \sum_{i=1}^{w} \beta_{i} \cdot \bar{\mathrm{h}}^{i}
\end{aligned}
\end{equation}
where $\alpha_{i}$ and $\beta_{i}$ are the weights for user representation and item representation,respectively. Finally, to jointly optimize parameters in embedding matrix and parameters in controllers, it introduces a variant of DARTS that leverage the first order approximation as equation \ref{equ:first-orderapp}.

\textbf{AutoDim} \cite{zhao2021autodim} extends the AutoEmb to the field-aware embedding dimension search, which aims to automatically assign embedding dimensions for different feature fields instead of users/items. In particular, similar to AutoEmb, each feature field is defined in the $w$ embedding space and then unified and normalized into the same space. The major difference is that it utilizes the Gumbel-softmax operation \cite{jang2016categorical} with architecture weights to select embedding dimensions, which could also deal with the end-to-end indifferentiable issue due to the hard dimension selection. Finally, the DARTS-based optimization algorithm is also employed to optimize the embedding matrix and the architectural weights jointly.

\begin{table*}[htbp]
  \centering
%   \tiny
  \caption{Summary of Auto-EDS methods.}
  \begin{adjustbox}{max width=\textwidth}

\begin{tabular}{cccccc}
\toprule
\textbf{Categories} & \textbf{Methods} & \textbf{Search Space} & \textbf{Search Strategies} & \textbf{Search Levels} & \textbf{Tasks} \\
\midrule
\midrule
Heuristic & MDE (ISIT'2021) \cite{ginart2021mixed} & ---   & Human Rules & Feature Fields & CTR \\
\midrule
\multirow{5}[10]{*}{HPO} & NIS (KDD'2020) \cite{joglekar2020neural} & $(S+1)^T$ & ENAS-based & Features & Top-K Recommendation \& CTR \\
\cmidrule{2-6}      & DNIS (ArXiv'2020) \cite{cheng2020differentiable} & $2^{Sd}$ & DARTS-based & Features & Rating Prediction \& Top-K Recommendation \& CTR \\
\cmidrule{2-6}      & AutoEmb (ICDM'2021) \cite{zhao2020autoemb} & $2^{(U+I)w}$ & DARTS-based & Users/Items & Rating Prediction \\
\cmidrule{2-6}      & AutoDim (WWW'2021) \cite{zhao2021autodim} & $w^{F}$ & DART-based & Feature Fields & CTR \\
\cmidrule{2-6}      & ESAPN (SIGIR'2020) \cite{liu2020automated} & $2^{(U+I)w}$ & ENAS-based & Users/Items & CTR \& Rating Prediction \\
\midrule
\multirow{5}[10]{*}{Embedding Prunning} & PEP (ICLR'2021) \cite{liu2021learnable} & ---   & Soft Threshold Reparameterization & Features & CTR \\
\cmidrule{2-6}      & AMTL (CIKM'2021) \cite{yan2021learning} & ---   & Adaptively-Masked Layer & Features & CTR \\
\cmidrule{2-6}      & RULE (KDD'2021) \cite{chen2021learning} & $(2^{n_1}-1)^{ \frac{| \mathcal{I} |}{n_2}}$ & Evolutionary Search & Items & Top-K Recommendation \\
\cmidrule{2-6}      & DeepLight (WSDM'2021) \cite{deng2021deeplight} & ---   & Greedy Algorithm & Features/DNN & CTR \\
\cmidrule{2-6}      & SSEDS (SIGIR'2022) \cite{qu2022single} & ---   & Dimension Salience Criterion & Feature Fields & CTR \\
\bottomrule
\end{tabular}%

    \end{adjustbox}
    \label{table:autoeds}
\end{table*}

\textbf{ESAPN} \cite{liu2020automated} aims to search embedding dimensions for users and items dynamically based on their popularity by an automated reinforcement learning agent. However, unlike AutoEmb using a soft selection strategy, ESAPN performs a hard selection on candidate embedding dimensions, which could effectively reduce the storage space. Specifically, it consists of a deep recommendation model performing personalized recommendations and two policy networks learning embedding dimensions for users and items from a discrete candidate dimension set. The deep recommendation component is similar to AutoEmb which the embeddings of users/items are defined in $w$ various embedding spaces and then are transformed into the largest dimension via linear transformations and batch normalization. For policy networks which are multi-layer perceptrons with multiple hidden layers, they take the frequency and current embedding dimension of users/items as inputs (i.e., states for reinforcement learning agent) and output two possible actions: enlarging the current dimension to the next larger dimension or unchanging the current dimension. The design of such actions is because they assume that users/items with a higher frequency have
larger embedding dimensions. Furthermore, the reward of policy networks is defined based on the difference between the current prediction loss and the previous losses. Finally, inspired by ENAS \cite{pmlr-v80-pham18a}, the recommendation model and policy networks are optimized in an alternative fashion, which optimizes policy networks using the sampled validation data and uses training data to optimize the recommendation model.

Although HPO-based Auto-EDS methods could effectively learn MD embeddings in different levels (i.e., features, feature fields, and users/items), such kinds of methods still suffer from several issues. (1) \textbf{Resources consumption}: to maintain embeddings on different embedding spaces \cite{zhao2020autoemb,zhao2021autodim,liu2020automated}, they have to maintain additional embedding matrixes with different dimensions, which consumes a huge amount of storage space. (2) \textbf{Expensive optimization}: the model overall model optimization is time-consuming due to optimizing the extra parameters in controllers \cite{liu2020automated,joglekar2020neural}. (3) \textbf{Severe assumption}: the assumption that the high-frequency features (users/items) should be assigned with the larger embedding dimensions \cite{liu2020automated,joglekar2020neural} might not always be satisfied due to the complex situations in recommendations.

\subsection{Embedding Pruning Methods}
Another research line in this field considers the Auto-EDS as the embedding pruning problem, which performs embedding pruning over the whole embedding matrix using different pruning strategies such that the MD embeddings are automatically obtained. Thus, the key idea of such kinds of methods is to build memory-efficient models by identifying and removing the redundant parameters in the embedding matrix and keeping the recommendation as accurate as possible.

For example, \textbf{PEP} \cite{liu2021learnable} introduces the learnable thresholds to identify the importance of parameters in the embedding matrix. In particular, inspired by Soft Threshold Reparameterization \cite{kusupati2020soft}, it directly performs adaptively pruning as follows:
\begin{equation}
    \hat{\mathbf{E}} = \mathcal{S}(\mathbf{E},s) =  sign(\mathbf{E})ReLU(abs(\mathbf{E})-\frac{1}{1+e^{-s}})
    \label{equ:pep}
\end{equation}
where $\hat{\mathbf{E}}$ and $\mathbf{E}$ are the re-parameterized embedding matrix and original embedding matrix, respectively. $abs(\cdot)$ is the absolute operation. $s$ is learnable threshold(s) that could be updated by gradient descent, and $sign(\cdot)$ is sign function. Furthermore, the sub-gradient is utilized to solve the non-differentiability of equation (\ref{equ:pep}) as below:
\begin{equation}
\mathbf{E}^{(t+1)} \leftarrow \mathbf{E}^{(t)}-\eta_{t} \nabla_{\mathcal{S}(\mathbf{E}, s)} \mathcal{L}\left(\mathcal{S}\left(\mathbf{E}^{(t)}, s\right), \mathcal{D}\right) \circ \mathbb{L}\left\{\mathcal{S}\left(\mathbf{E}^{(t)}, s\right) \neq 0\right\}
\end{equation}
where $\eta_{t}$ denotes $t$-th step learning rate, and $\circ$ is element-wise product. $\mathcal{L}$ and $\mathcal{D}$ are the loss function (e.g., the cross-entropy loss) and the dataset, respectively. $\mathbb{L}\{\cdot\}$ is the indicator function. In this way, the threshold(s) $s$ and embedding matrix $\mathbf{E}$ could be jointly trained by gradient descent. After that, it could mask those dropped parameters in $\mathbf{E}$ to obtain a pruned embedding matrix which could be utilized to re-train the based model according to the Lottery Tickey Hypothesis \cite{frankle2018lottery}.

\textbf{ATML} \cite{yan2021learning} proposes to use an Adaptively-Masked Twins-based layer (i.e., AMTL) behind the original embedding layer to learn a mask vector which is utilized to mask those redundant dimensions in the embedding matrix. Specifically, to leverage the feature frequency knowledge, the AMTL takes the feature frequency vectors as input and then uses two branches (i.e., h-AML and l-AML) to handle high-frequency and low-frequency samples, respectively, so that the parameters in l-AML will not be dominated by the high-frequency samples. Moreover, it introduces a soft decision strategy to determine the high- or low- frequency samples, which uses a weighted sum of outputs of h-AML and l-AML as follows:
\begin{equation}
\begin{aligned}
output_{L}^{(A M T L)} &=\alpha_{i} * output_{L}^{(h-A M L)}+\left(1-\alpha_{i}\right) * output_{L}^{(l-A M L)} \\
\alpha_{i} &=\operatorname{sigmoid}\left(\operatorname{norm}\left(q_{i}\right)\right)
\end{aligned}
\end{equation}
where $output_{L}^{(h-AML)}$ and $output_{L}^{(l-AML)}$ are the $L$-th outputs of l-AML and h-AML, respectively, and $\alpha_{i}$ is the weight which is influenced by the feature frequency $f_{i}$. Then, a temperated softmax function \cite{hinton2015distilling} is applied on $output_{L}^{AMTL}$ to obtain the probability to select different embedding dimensions. 

% Deeplight \cite{deng2021deeplight} proposes to prune both parameters in the embedding layer and DNN layer to solve the high-latency issues in CTR prediction. 

\textbf{RULE} \cite{chen2021learning} proposes an on-device recommendation paradigm with elastic embeddings dealing with various memory budgets for devices. The paradigm is divided into learning full embeddings and searching elastic embeddings for items with an evolutionary algorithm. As argued by the authors, the learning time of embedding is far beyond the searching time, leading to a once-for-all paradigm, which adapts the learned embedding table to local on-device recommenders with heterogeneous resource budgets. In the learning phase, Bayesian personalized ranking (BPR) \cite{rendle2012bpr} loss is applied to optimize the full embedding and regularization terms maintaining the diversity of the learned embedding blocks, benefiting the subsequent search procedure. In the deployment phase, the evolutionary algorithm \cite{real2019regularized} with a single-layer feed-forward network estimator is adopted to search the desired embedding blocks under the memory budgets. The search space is $(2^{n_1}-1)^{ \frac{| \mathcal{I} |}{n_2}}$, where the hyper-parameters $n_1$, $n_2$ represent the number of embedding blocks for an item, and the number of item groups respectively. The embedding pruning is done by maintaining the searched embeddings by evolutionary algorithms.

\textbf{Deeplight} \cite{deng2021deeplight} proposes to prune both parameters in the embedding layer and DNN layer to solve the high-latency issues in CTR prediction. The weight matrices of the DNN component are pruned to remove the connections. Thus, the sparse DNN component with less computation complexity contributes to the training acceleration. The field pair interaction matrix is pruned as field pair selection. Moreover, the elements in the embedding vectors are pruned to be sparse embedding vectors. Considering the majority of the parameters in deep learning models for click prediction are feature embeddings, Deeplight is classified into Auto-EDS in our taxonomy. The model is trained for a few epochs, and weights with the smallest values are removed based on adaptive sparse rate. The search strategy can be summarized as the variant of the greedy algorithm with weak sub-modular optimization \cite{das2011submodular}.

\textbf{SSEDS} \cite{qu2022single} proposes a single-shot embedding pruning method named SSEDS. In particular, it first pre-trains a traditional CTR model with unified embedding dimensions, and then utilizes the proposed criterion which could measure the importance of embedding dimensions only in one forward-backward pass to obtain the salience scores of each dimension. In this way, the redundant dimensions could be removed based on the dimension importance ranking and the parameter budget. Furthermore, since the obtained mixed-dimensional embeddings could not be directly applied to traditional CTR models due to some feature interaction operations (e.g., the dot product) requiring all embeddings with the same dimension, SSEDS utilizes the additional transform matrices to align all dimensions and re-trains the slim model.

Although the embedding pruning-based Auto-EDS methods could build the memory-efficient model via selectively reducing parameters in the embedding matrix, these methods generally require an iterative optimization procedure for both parameters in the embedding matrix and additional parameters used to prune, which is time-consuming.

\section{Automated Feature Interaction Search (Auto-FIS)}
\label{sec:autofis}
% In this section, we collect papers related to AutoML for deep Recommender Systems. Based on how these works automatically alter the component of the deep Recommender Systems, we group them into four categories, embedding size search methods, feature interaction search methods, loss search methods and model search methods. 
\begin{figure*} 
\centering
\includegraphics[width=\textwidth]{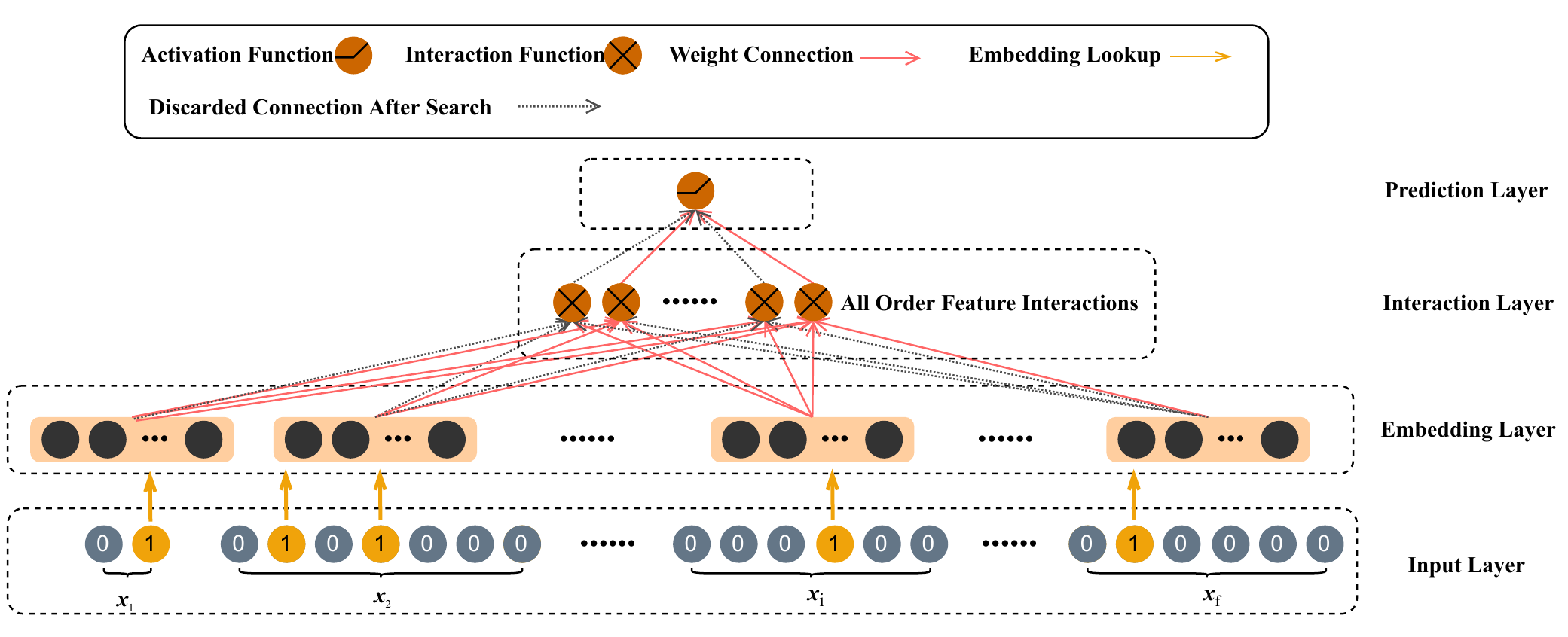}
\caption{The framework of Auto-FIS.}
\label{fig:autofis}
\end{figure*}

Feature interactions combine individual features of users, items, and other context information. For instance, the user's age and generic of the downloaded application can be combined together as the $2^{nd}$ order feature interactions and contribute to the users' preference prediction in the application recommendation scenario. The order implements the number of combined features. The effectiveness of $2^{nd}$ order feature interactions has been proved in Factorization Machines (FMs) \cite{rendle2010factorization}  and their variants \cite{pasricha2018translation}\cite{juan2016field}.  Meanwhile, the high-order ($p^{th} $ order with $p \geq 3$) feature interactions is approximated by Higher-order Factorization Machine (HOFM) \cite{blondel2016higher}. 

With the prosperity of deep neural networks, many deep learning recommender models have been proposed due to their better performance than traditional models. Product-based Neural Network (PNN) \cite{qu2016product} extracts the high-order feature interactions and discards the lower ones. Wide\&Deep \cite{cheng2016wide} and DeepFM \cite{guo2017deepfm} learn the low-order and high-order feature interactions by a shallow component and a deep component and state that both the low-order and high order feature interactions play a significant role in context-aware recommender systems.

However, enumerating all high-order  feature interactions is time and space-consuming. When there are $f$ features in total, the number of $p^{th}$ order interaction terms is ${f}\choose{p}$. Even for $2^{nd}$ order feature interactions, simply listing all combinations may introduce useless interactions as noise and disturb the model performance. Therefore, how to keep the necessary feature interactions and filter out the useless ones arouse people's interest. We summarize these methods in Table~\ref{table:autofis}. Many automated feature interaction search methods are proposed to deal with the following three challenges mainly. (1) The desired beneficial feature interaction sets are discrete.
(2) Both lower and higher order feature interactions are highly correlated. (3) The priorities of the low- and high-order interactions in the search procedure should be considered. Concretely, as shown in Fig. \ref{fig:autofis}, we introduce an overview of Automated Feature Interaction Search (Auto-FIS) architecture. The key components are the set containing all possible order feature interactions and the discarded connections, which are determined by the search procedure.

Sparse Factorization Machines (SFMs) \cite{zhao2017meta} determine the relevant user features and item features based on their contributions to the prediction model by sparsity regularization \cite{tibshirani1996regression}. If one feature does not contribute to the predictive modeling, entire feature interactions related to it will be deactivated. If some significant high-order feature interactions play a role in the prediction only as a whole rather than individuals, SFMs may discard them even it has a recovery procedure to cover features that are relevant to user-item prediction. Moreover, the feature interaction selection strategy of SFMs is identical for all users. Same interactions may play a more important role to one user than others. The most intuitive solution is to build distant SFMs for every user, which does not preserve the benefit of collaborative filtering. 

Thus, Bayesian Personalized Feature Interaction Selection (\textbf{BP-FIS}) \cite{chen2019bayesian} is proposed to adaptively select interactions for individual users by Bayesian Variable Selection \cite{mitchell1988bayesian}. The prediction is modified as below:

\begin{equation}
\hat{y} =  \sum^{f}_{i=1} s_{ui} w_{i}  \mathbf{x}_i +   \sum^{m}_{i=1}\sum^{m}_{j>i} s_{uij} w_{ij}  < \mathbf{e}_i, \mathbf{e}_j>  \mathbf{x}_i  \mathbf{x}_j + b_u
\end{equation}
where $b_u$ is the bias for user $u$. The single feature interaction weights $\mathcal{W} = \{w_{i}\} \cup \{w_{ij}\}$ are learned for all users, while personalized feature interaction selection variables $\mathcal{S} = \{ s_{ui}\} \cup \{ s_{uij} \}$ indicate the $1^{st}$ order and $2^{nd}$ order feature interaction selection for individual user $u$. $|\mathcal{U}|$ is the number of users. Due to the immense search space $O(|\mathcal{U}| \cdot F^2)$, BP-FIS proposes Hereditary Spike-and-Slab Prior (HSSP) based on Spike-and-Slab Priors (SSPs) \cite{bernardo2003bayesian} to attain the heredity property of feature interaction.
Strong heredity indicates that if the $1^{st}$ order interactions $ \mathbf{x}_i$ and $ \mathbf{x}_j$ are chosen, their combination, i.e., the $2^{nd}$ order interaction $< \mathbf{e}_i, \mathbf{e}_j>$ would be selected, while weak heredity indicates the selection possibility of their combination to be $\pi_2$ when only one of the $1^{st}$ order interactions is selected. To be specific, the HSSP is stated as:
% \begin{equation}

% \end{equation}
\begin{align}
\label{eqn:eqlabel}
\begin{split}
  p(s_{ui} = 1) = p(s_{uj} = 1 ) &= \pi_1 \\
            p(s_{uij} | s_{ui}s_{uj} = 1) &= 1 \ \ (strong \ heredity)\\
            p(s_{uij} = 1 | s_{ui} + s_{uj} = 1)  &= \pi_2 \ (weak \ heredity) \\ 
            P(s_{uij} = 1 | s_{ui}+s_uj =0) &= 0
\end{split}
\end{align}
where $\pi_1, \pi_2 \in \{0, 1\}$ are constant values. Inspired by Variational Auto-Encoder (VAE) \cite{kingma2013auto}, Stochastic Gradient Variational Bayes (SGVB) estimator is proposed to approximate posteriors of the latent variables and optimize the model. BP-FIS can be combined with both linear FM and neural FM.

\textbf{AutoFIS} \cite{liu2020autofis} is proposed to learn the feature interactions in recommender system models by adding an attention gate to every potential feature interaction, rather than simply enumerating all the two-order feature interactions like Factorization Machines. There are two stages for AutoFIS: the search stage and the re-train stage.
% for click-through rate (CTR) and conversion rate (CVR) aimed task.

In the search stage, equation \ref{eqau:autofis} shows the interaction layer of the factorization model in AutoFIS.
\begin{equation}
\boldsymbol{l}_{AutoFIS} = <\mathbf{w},\mathbf{x}> + \sum^{f}_{i=1}\sum^{f}_{j>i} \alpha_{(i,j)}  <\mathbf{e}_i,\mathbf{e}_j>
\label{eqau:autofis}
\end{equation}

% dd
Instead of searching the desired discrete combination set from $O(2^{F^2})$ search space for two-order feature interactions, the wight $\alpha_{(i,j)}$ represents the importance of the interaction between vector $e_i$ and vector $e_j$, and determinates whether preserve or delete this interaction. After feeding the output of the interaction layer as the input of multi-layer perceptron (MLP), the user's preference for item can be calculated via:
\begin{equation}
\hat{y} = sigmoid(MLP(\boldsymbol{l}_{AutoFIS}))
\end{equation}

Architecture parameters $\boldsymbol{\alpha} = \{\alpha_{(i,j)}\}_{i=1,\cdot\cdot\cdot,m,i < j \leq f}$ reveal the contribution of every feature interaction towards the final prediction. Furthermore, the architecture parameters $\boldsymbol{\alpha} $ are trained by generalized regularized dual averaging (GRDA) optimizer \cite{chao2019generalization}, while other parameters (such as  $\mathbf{w}$ in equation \ref{eqau:autofis} ) are updated by Adam optimizer \cite{kingma2014adam}. These two optimizations are conducted jointly in one gradient descent step, unlike the bi-level optimization algorithm in DARTS \cite{liu2018darts}.

In the re-train stage, architecture parameters $\boldsymbol{\alpha}$ are fixed as $\boldsymbol{\alpha}^*$ after search stage. Feature interactions that benefit the final prediction have higher $\alpha^*_{(i,j)}$ values. The feature interaction layer in equation \ref{eqau:autofis} is modified as:

\begin{equation}
l_{AutoFIS} = <\mathbf{w},\mathbf{x}> + \sum^{m}_{i=1}\sum^{m}_{j>i} \alpha_{(i,j)^*} \gamma_{(i,j)}  <\mathbf{e}_i,\mathbf{e}_j>
\end{equation}
where $\gamma_{(i,j)}$ depicting the gate status is set as 0 when  $\boldsymbol{\alpha}^*$ = 0, otherwise 1. With attention unit $\boldsymbol{\alpha}^*$, the model is further trained by relative important interactions, and all the parameters are learned by Adam optimizer \cite{kingma2014adam}.

Unlike AutoFIS simply focusing on two-order feature interaction selections, \textbf{AutoGroup} \cite{liu2020autogroup} considers the high-order feature interaction search as a structural optimization problem to identify the useful high-order interactions. The search space is $O(2^{f^{K}})$, where pre-defined $K$ represents the maximum order of feature interactions. Firstly, for $p^{th}$ order of feature interactions, it selects $n_p$ feature sets. $\mathcal{F}_j^p$ represents the $j^{th}$ feature set for $p^{th}$ order. Every feature $\mathbf{f_i}$ is possible to be included in  $\mathcal{F}_j^p$ indicated by the structural parameter $\alpha^{p}_{i,j}$. 

After the automatic feature grouping stage, the representation of a feature set $\mathcal{F}_j^p$ is defined as the weighted sum of feature embedding within it: 
\begin{equation}
\mathbf{g}^p_j = \sum_{\mathbf{f}_i \in \mathcal{F}_j^p} w_i^p \mathbf{e}_i
\end{equation}
where $\mathbf{e}_i$ is the embedding of feature $\mathbf{f}_i$ and $w_i^p$ is trainable weight parameter. Inspired by FM \cite{rendle2010factorization}, the time complexity is reduced from $O(F^p)$ to $O(F)$, owning to feature interaction calculation method within one feature set $\mathcal{F}_j^p$.

\begin{equation}
  I_{j}^{p} =
    \begin{cases}
      (\mathbf{g}^p_j)^p - \sum_{f_i \in \mathcal{F}_j^p}( w_i^p \mathbf{e}_i) ^p , & \text{$p \geq 2 $}\\
        \mathbf{g}^p_j, & \text{$p = 1 $}\\
      
    \end{cases}       
\end{equation}
where $(\mathbf{g}^p_j)^p$ represents the sum of all the embedding components of the embedding generated by $p$ times the element-wise product of $\mathbf{g}^p_j$ with itself.
There are $\sum_{i=1}^K n_i$ interaction results in total. All the interactions are concatenated and fed into an MLP. The prediction is calculated by:

\begin{equation}
\hat{y} = sigmoid(MLP(concat(\underbrace{I^1_1,\cdot\cdot\cdot,I^1_{n_1}}_\text{$1^{st}$ order},\cdot\cdot\cdot,\underbrace{I^K_1,\cdot\cdot\cdot,I^K_{n_K}}_\text{$K^{th}$ order}))) 
\end{equation}
% $\hat{y} = sigmod(MLP(concat()))$.

AutoGroup optimizes the structural parameters and network weights (e.g., embedding parameters and MLP parameters) alternatively by gradient descent, similar to DARTS \cite{liu2018darts}, since two kinds of trainable parameters are highly dependent on each other.

\begin{table*}[htbp]
  \centering
  \caption{Summary of Auto-FIS methods.}
  \begin{adjustbox}{max width=\textwidth}
% Table generated by Excel2LaTeX from sheet 'Sheet1'
\begin{tabular}{ccccc}
\toprule
\textbf{Methods} & \textbf{Search Space} & \textbf{Search Strategies} & \textbf{Order} & \textbf{Tasks} \\
\midrule
\midrule
BP-FIS (SIGIR'2019) \cite{chen2019bayesian} & $O(|\mathcal{U}| \cdot F^2)$ & Bayesian Optimization & Explicit Low Order & Top-K Recommendation  \\
\midrule
AutoFIS (KDD'2020) \cite{liu2020autofis} & $O(2^{F^2})$ & Gradient-based & Explicit Low Order & CTR \\
\midrule
AutoGroup (SIGIR'2020) \cite{liu2020autogroup} & $O(2^{F^k})$ & Gradient-based & Explicit High Order & CTR  \\
\midrule
FIVES (KDD'2021) \cite{xie2021fives}  & $O(2^{KF^2})$  & Gradient-based & Explicit High Order & CTR  \\
\midrule
FROFIT (NeurIPS'2021) \cite{gao2021progressive} & $O(RFK)$ & Gradient-based & Explicit High Order & CTR \\
\midrule
$L_0$-SIGN (AAAI'2021) \cite{su2021detecting} & $O(2^{F^2})$ & MLP   & Explicit Low Order & CTR \\
\midrule
HIRS (KDD '2022) \cite{su2022detecting} & $O(n_1 \cdot 2^{F})$ & MLP   & Explicit High Order & Top-K Recommendation  \\
\bottomrule
\end{tabular}%
    \end{adjustbox}
  \label{table:autofis}%
\end{table*}%

With the prosperity of Graph Neural Network (GNN), \textbf{$L_0$-SIGN} \cite{su2021detecting} implements GNN techniques to tackle the feature interaction search problem. Given a Graph $\mathcal{G} = (\mathcal{N},\mathcal{E})$, where $n_i \in \mathcal{N} $ represents feature $\mathbf{f}_i$, edge $e_{i,j} \in \{1,0\}$ from the edge set $\mathcal{E} = \{e_{i,j}\}_{i,j = 1,\cdot\cdot\cdot,f}$ represents whether select the feature interaction between feature $\mathbf{f}_i$ and feature $\mathbf{f}_j$. Initially, the edge set is empty $\mathcal{E} = \emptyset$, and the task of searching for beneficial feature interactions is converted to the edge prediction in graph $\mathcal{G}$. Therefore, the complexity of the search space is $O(2^{F^2})$. $L_0$-SIGN implements an MLP to predict the edge:
\begin{equation}
e_{i,j}^{\prime} = MLP(\mathbf{e}_i \circ \mathbf{e}_j)
\end{equation}
where the feature embedding $\mathbf{e}_i$ and $\mathbf{e}_j$ are obtained by equation \ref{equa:embedding1} and \ref{equa:embedding2}. Set $\mathcal{E}^{\prime}$ including all detected edges $e_{i,j}^{\prime}$, is performed with an $L_0$ activation regularization to minimize the  number of searched beneficial interactions. After determining the edge set $e_{i,j}^{\prime}$, the embedding  $\mathbf{e}_i^{(1)} = \mathbf{e}_i$ is updated iteratively $t$ times by a linear aggregation function $\psi(\cdot)$ (e.g. element-wise summation/mean):
\begin{equation}
(\mathbf{e}_i)^{t} = \psi(s_i^{(t-1)})
\end{equation}
where $t$ is the iteration index, and $s_i^{(t-1)}$ is the set of statistical interaction analysis outcomes between  $\mathbf{e}_i^{(t-1)}$ (i.e. the embedding of node $i$ at ${(t-1)}^{th}$ iteration) and embeddings of its neighbours. The final prediction $\hat{y}$ is calculated as:

\begin{equation}
\hat{y} = \frac{\sum_{i=1}^{f}g(\mathbf{e}_i^{(t)})}{f} + b
\end{equation}
where the linear function $g(\cdot)$ (e.g., weighted sum function) converts the node embedding to a scalar value, and $b$ is the bias term.

While $L_0$-SIGN can only search for the $2^{nd}$-order feature interaction, \textbf{HIRS} \cite{su2021detecting} extends it to arbitrary order of feature interaction by introducing the hyper-edge set $\{edge_j\}^{n_1}_{j=1}$, where hyper-parameter $n_1$ represents the number of intended search feature interaction. Each hyper-edge is $F$-dimensional binary vector $edge^i_j \in \{0, 1\}$, where  $edge^i_j = 1$ represents that the $j^{th}$ hyper-edge links to feature $\mathbf{f}_i$. The hyper-edge prediction module contains a multi-layer perceptron, looking through the search space with complexity $O(n_1 \cdot 2^{F})$.

Under the same settings of graph $\mathcal{G}$, \textbf{FIVES} \cite{xie2021fives} extends the interaction search to high-order with a graph neural network and an adjacency tensor. Adjacency tensor $\mathbf{A} \in \{0,1\}^{K \times F \times F}$ indicates the interactions at every order $k \leq K$. $\mathbf{e}_i^{(k)}$ is the representation for node $i$ at order $k$, and $\mathbf{e}_i^{(1)}= \mathbf{e}_i$. Based on the proposition that the interaction of uninformative ones is unlikely to build an informative feature, when $\{\mathbf{A}_{i,j}^{(k)}\}_{i,j=1,\cdot\cdot\cdot,f}$ is active,  $\mathbf{e}_i^{(k)}$ is calculated as:
\begin{equation}
\mathbf{e}_i^{(k)} = (MEAN_{j|A_{i,j}^{(k)}=1} \{ \boldsymbol{W}_j \mathbf{e}_j^{(1)}\}) \circ \mathbf{e}_i^{(k-1)}  
\end{equation}
where "MEAN" is the aggregator, $\circ$ is the element-wise product, and $\mathbf{W}_j$ represents the transformation matrix for node $j$. Under the assumption that $(\mathbf{W}_j n_j^{(1)}) \circ n_i^{(1)}$ can express feature interaction $\mathbf{f}_i \otimes \mathbf{f}_j$, the $k^{th}$ node representation $\mathbf{e}^{(k)} = [\mathbf{e}^{(k)}_1,\cdot\cdot\cdot\mathbf{e}^{(k)}_f]$ matches the $k^{th}$ order feature interactions, and adjacency tensor $\mathbf{A}$ determines which features to be selected. Thus, the search space is $O(2^{K F^{2}})$. 

The search problem of FIVES is a Bi-level optimization where $\mathbf{A}$ is the upper-level variable, and other model variables are the lower-level ones,

% $\mathcal{X} = \{X_i\}_{i = 1,\cdot\cdot\cdot,N}$ represents the set of user-item examples 

Following the settings in AutoFIS, where gradient descent NAS method with relaxation turns the search space into symmetric parameter matrix $\boldsymbol{W} \in \mathbb{R}^{F \times F}$, \textbf{PROFIT} \cite{gao2021progressive} finds that parameter matrix $\boldsymbol{W}$ has several dominant singular values, and exhibits a low-rank property. Therefore, PROFIT proposes a distilled search space $\boldsymbol{A}$ by symmetric CP decomposition \cite{kolda2009tensor}, extending to $k^{th}$ order interactions:

\begin{equation}
\boldsymbol{A} = \sum^R_{r=1} \underbrace{\boldsymbol{\beta}^{r} \circ \cdot \cdot \cdot \circ \boldsymbol{\beta}^{r}}_\text{repeat K times} 
\end{equation}
where approximated vector $\boldsymbol{\beta}^r \in \mathbb{R}^{1 \times F}$ is updated by gradient descent, and the distilled search space is based on low-rank approximation with a positive integer $R \ll F$. The authors state that the complexity is $O(RFK)$. Following the idea that different orders of feature interactions are highly correlated, PROFIT proposes a progressive gradient descent to learn the high-order after the low one. Specifically, when learning the $k^{th}$ order interaction, the architecture parameters $\{\boldsymbol{\beta^{i}}\}_{i-1,...,k-1}$ are fixed.

BP-FIS has two major drawbacks. (1) BP-FIS limits the interactions up to two orders. (2) Selection for every user may be a waste of resources. Group-level penalization could be an efficient manner to control the size of the search space.

AutoFIS, AutoGroup, and AutoCross adopt NAS on the continuous search space with the help of  continuous relaxation. The search problem is modified from choosing one interaction to calculating the weight of that interaction. The complexity of search space is reduced from the original search space to the number of the weight parameters $\sum^{K}_{i=1} {F \choose i}$. However, they still encounter several issues. (1) \textbf{Slow optimization procedure:} The number of weight parameters is significantly greater than existing gradient-based NAS methods \cite{liu2018darts,xie2018snas}, which usually have less than one hundred parameters. (2) \textbf{Ignorance on the order-priority property:} They neglect the relations between different order interactions. 

Both implementing GNN techniques to search for beneficial feature interactions, $L_0$-SIGN focuses on $2^{nd}$ order feature interactions. In contrast, FIVES extends to high-order under the proposition that it is improbable to produce the instructive feature interactions from the uninformative interactions, which may miss particular significant interactions in some scenarios. Although the search space of PROFIT has been dramatically reduced, the choice of hyper-parameter $R$ influences the final performance.

% Table generated by Excel2LaTeX from sheet 'Sheet1'
\begin{table*}[htbp]
  \centering
  
  \caption{Summary of Auto-MAS methods.}
  \begin{adjustbox}{max width=\textwidth}
\begin{tabular}{cccc}
\toprule
\textbf{Methods} & \textbf{Search Space} & \textbf{Search Strategies} & \textbf{Tasks} \\
\midrule
\midrule
AutoCTR (KDD'2020) \cite{song2020towards} & $O(4^{n} \cdot 2^{\frac{n (n-1)}{2}})$ & Multi-Objective Evo & CTR \\
\midrule
AMEIR (IJCAI'2021) \cite{zhaoameir} & $O(32^{n_1}\cdot2^{F^k}\cdot 40^{n_2})$ & One-shot Random Search & CTR \\
\midrule
AutoIAS (CIKM'2021) \cite{wei2021autoias} & $O(F^{n_1}\cdot (2^{F^2})^{ n_2 + n_3 + 1 + n_5}\cdot n_6)$ & MLP   & CTR \\
\midrule
{NASR (WWW'2022) \cite{cheng2022towards}} & $4^{n_1}$ & Greedy Search & Top-K Recommendation  \\
\bottomrule
\end{tabular}%

    \end{adjustbox}
  \label{table:automas}%
\end{table*}%
\section{Automated Model Architecture Search (Auto-MAS)}
\label{sec:automas}
Numerous classical approaches \cite{lian2018xdeepfm,wang2017deep} have demonstrated the significance and effectiveness of feature interactions in coping with high-cardinality feature attributes and large-scale datasets. They discover the explicit low-order feature interactions and combine them with explicit or implicit high-order feature interactions. For instance, Wide\&Deep \cite{cheng2016wide} and DeepFM \cite{guo2017deepfm} learn the explicit low-order and implicit high-order feature interactions by a shallow and a deep component. The section on feature interaction search methods systematically reviews various techniques to identify beneficial feature interactions. However, existing methods search for one particular layer and leave other components of the deep recommender systems hand-crafted, causing three problems. (1) An integrated recommender system model cannot be directly obtained by the above-mentioned automated search methods, and domain experts are needed to design other components manually. (2) An whole well-performed architecture is less likely to be acquired due to other hand-crafted components, even the searched one finding the best candidate. (3) The generalization capacity is decreased by the limited search space for one specific layer rather than the whole architecture. 

Moreover, the design of MLP layers in classical approaches may not be optimal both in efficiency and effectiveness. Diamond MLP frameworks may outperform rectangular, and triangle frameworks \cite{zhang2016deep}, which aroused people's interest in the MLP design. Experts cannot attempt all the potential design architecture. Therefore, automated model architecture search methods are employed to mitigate human efforts and search for an automatically designed task-specific architecture that organically combines informative embeddings and various feature interactions. There are mainly three challenges. (1) \textbf{Accurate search space:} The search space of automated model architecture search methods should be carefully designed, which includes popular and effective human-crafted architectures. In the meantime, the search space cannot be exceedingly general. (2) \textbf{Search Efficiency:} Automated model architecture search methods should be efficient, especially for the design of the recommender system architecture. A practical model encounters a billion-level of user-item data in the industry, for example, the private dataset in AutoFIS \cite{liu2020autofis} from Huawei App Store. (3) \textbf{Ability to distinguish:} Recommender systems with diverse architecture may lead to similar performance on the validation dataset since minor improvements to the experiment contribute to significant refinements in practice. Automated model architecture search methods should be sensitive to slight progress. We summarize these methods in Table~\ref{table:automas}.

\textbf{AutoCTR} \cite{song2020towards} proposes a two-level hierarchical search space, which includes the popular human-crafted deep recommender system architectures by abstracting different feature interaction methods into virtual blocks. The inner search space contains the choice of virtual blocks and detailed hyperparameters inside each of those blocks. Based on the existing literature, the virtual blocks can be selected from FM, MLP, and dot products. Instead of simply stacking the blocks sequentially, AutoCTR implements a direct acyclic graph (DAG) to connect different feature interaction blocks organically, rather than directly stacking different components sequentially. A block may receive outputs from any preceding block, including numerous unexplored architectures. The outer search space contains all the possible block combination choices. Let $n$ represent the number of virtual blocks, controlling the complexity of the architectures. There are four basic candidate operators for blocks, and $\frac{n (n-1)}{2}$ possible connections for $n$ blocks. Therefore the search space is $O(4^{n} \cdot 2^{\frac{n (n-1)}{2}})$.

A multi-objective evolutionary search is used to explore the two-level hierarchical search space. The first procedure is survivor selection. Only top-$p$ architectures survive according to the metric $g$:

\begin{equation}
g(q, a_A, r_A, c_A) = \mathbb{L}_{[a_{A} \leq q]} \cdot (\mu_{1}a_A+\mu_2 r_A + \mu_3 c_A)
\end{equation}

where $q$ is the hyperparameter larger than $p$, and the indicator function $\mathbb{L}(\cdot)$ filters out architectures older than q. Three objectives of the architecture $A$: age $a_A$, performance $r_A$, and complexity $c_A$ are balanced by parameters $\mu_1$, $\mu_2$, and $\mu_3$. To tackle the third challenge mentioned above, the probability of being selected as the parents in the second procedure is denoted as:

\begin{equation}
Prob(r_A) = \frac{{  r_A + \lambda -1 \choose \lambda}}{{  p+\lambda \choose 1+\lambda}}   \quad  \lambda \in \mathbb{N}^0
\end{equation}

where hyperparameter $\lambda$ balances exploitation and exploration. In the last procedure, a learning-to-rank strategy is adopted to guide the mutations by the gradient boosted tree learner and a pairwise ranking loss named LambdaRank \cite{burges2010ranknet}.

\textbf{AMEIR} \cite{zhaoameir} divides the recommendation models into three stages: behavior modeling, feature interaction, and MLP aggregation. Behavior modeling networks encapsulate users' specific and dynamic interests based on the sequential input features. Behavior modeling networks have three components: normalization, specific layer, and activation function, respectively selected from three candidate sets: \{layer normalization, None\}, \{convolutional, recurrent, pooling, attention\}, \{ReLU, GeLU, Swish, Identity\}. The search subspace of feature interactions is $O(2^{F^k})$, containing all possible combinations of $k^{th}$ order feature interactions. For the MLP aggregation stage, the number of hidden units has ten candidate values, and the activation function is chosen from \{ReLU, Swish, Identity, Dice\}. Therefore the complexity of the overall search is $O(32^{n_1}\cdot2^{F^k}\cdot 40^{n_2})$, where $n_1$ and $n_2$ represent the number of layers for behavior modeling and MLP respectively.

A one-shot random search is implemented to incorporate industrial requirements rather than a one-shot weight-sharing paradigm. It randomly samples child models from the search space and reserves the one with the best performance on the validation set. By gradually increasing the order of interactions, effective feature interactions are discovered. Beneficial interaction set with the fixed size is initialized with feature matrix $\mathbf{e}$ and updated by interacting with  $\mathbf{e}$. New interactions with the highest validation fitness are retained in the interaction set.

\textbf{AutoIAS} \cite{wei2021autoias} provides a more fine-grained search space including six components $\mathcal{S}_i$ with $n_i$ number of candidates, where $i \in \{1,\cdot\cdot\cdot,6\}$. First component $\mathcal{S}_1$ represents the embedding size for each binary feature vector $\{\mathbf{x}_i\}_{i=1,\cdot\cdot\cdot,f}$. $\mathcal{S}_2$ determines the unified projection embedding size before interaction for every $2^{nd}$ order interaction pair, and the third component $\mathcal{S}_3$ decides the feature interaction function. $\mathcal{S}_4$ inserts the result of one interaction $o(\mathbf{e}_i,\mathbf{e}_j)$ into the $l^{th}$ layer of MLP to mix the low- and high-order of feature interactions. The input of $l^{th}$ layer is the concatenation between the interaction result and the output of $(l-1)^{th}$ layer. The output of $l^{th}$ layer is calculated as:

\begin{equation}
\boldsymbol{h}_{l} = \sigma(concat[o(\mathbf{e}_i,\mathbf{e}_j),\boldsymbol{h}_{l-1}])
\end{equation}
where $l \in \{1,\cdot\cdot\cdot,L\}$. $\mathcal{S}_6$ decides the number of layers in MLP. AutoIAS implements architecture generator network to search through the immense search space with complexity $O(F^{n_1}\cdot (2^{F^2})^{ n_2 + n_3 + 1 + n_5}\cdot n_6)$. The Deep Neural Network (DNN) models the dependencies among different components by taking in previous components' states and generating the current component's selection probability on the candidate set. The performance prediction of a particular architecture is fastened by the ordinal parameter sharing on the supernet whose embedding size for different components is the largest one over the corresponding candidate set.

Unlike the above methods, \textbf{NASR} \cite{cheng2022towards} is an Auto-MAS model, searching hybrid architectures for the sequential recommendation. To alleviate the dilemma where increasing the number of depth layers leads to difficult training for neural networks \cite{he2016deep}, NASR adds one trainable parameter $\lambda_{L}$ to the residual connection:
\begin{equation}
\textbf{h}^{u}_{l+1} = \textbf{h}^u_{l} + \lambda_{l} \cdot Map_{l+1}((\textbf{h}^u_{l});  \mathcal{W}_{l+1})
\end{equation}
where $\textbf{h}^{u}_{l+1}$ and $\textbf{h}^u_{l}$ indicate the input and output for $l+1$-th residual block. $Map_{l+1}$ is the learnable residual mapping, and $\mathcal{W}_{l+1}$ represent all the parameters for $l+1$-th block. The trainable parameter $\lambda_{l}$ boosts convergence and improves sequential recommendation performance. The search space consists  of $n_1$ deep layers. For each layer, there are four candidates: two transformer block variants, and two temporal convolutional network variants. A greedy search strategy is utilized, accompanied by an unsupervised evaluation metric, which estimates the performance for each candidate layer.

One of the biggest challenges in automated model architecture search is the immense search space. All three methods suffer from this issue and propose distinct solutions to deal with it. AutoCTR implements multi-objective evolutionary search rather than gradient descent methods to parallelly explore the vast search space. AMEIR employs a one-shot random search to accelerate the search process, while AutoIAS utilizes DNN to model the dependencies among different components. Although the automated model architecture search is appealing and directly outputs well-performed recommender systems, simply stacking all the possible candidates leads to huge search space and is not practical in real-world applications. The inner relation between different components should be discovered, or some rule should be implemented to shrink the search space and serve as a search strategy guideline.

% Table generated by Excel2LaTeX from sheet 'Sheet1'
% Table generated by Excel2LaTeX from sheet 'Sheet1'
\begin{figure*} 
\centering
\includegraphics[width=\textwidth]{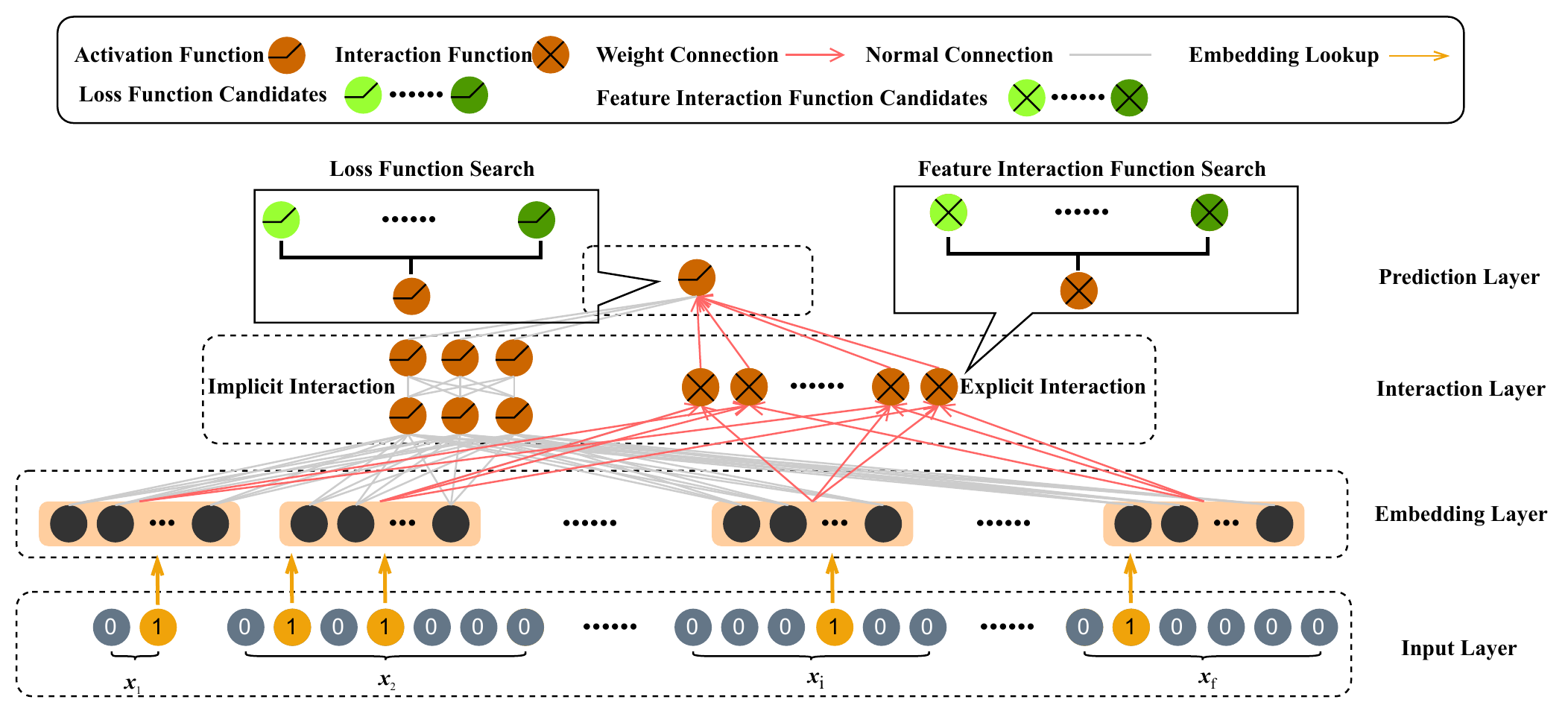}
\caption{The framework of Auto-OCS.}
\label{fig:autoocs}
\end{figure*}

\section{Automated Other Components Search (Auto-OCS)}
\label{sec:autoocs}
\subsection{Loss Function Search}
Deep neural networks (DNN) show promising results in recommender systems. One pivotal part of DNN training is the backpropagation, calculating the gradient based on the pre-defined loss function. However, choosing distinct loss functions is not universal and does not guarantee good performance. More appropriate gradients with a carefully-designed loss function may contribute to a better deep model.

People manually designed loss functions for specific tasks and purposes. A loss function with a higher value at the boundary position is proposed to improve the boundary metrics and satisfy the scenario when the boundary region is more significant \cite{qin2019basnet}. A large-margin softmax loss function in the image processing field is proposed to replace the common softmax loss function for desired feature discrimination learning \cite{liu2016large}. 

Despite the effectiveness of manually-designed loss functions, the exhausting design process requires human experts and a heavy workload. As shown in Fig.~\ref{fig:autoocs}, the automated loss search method selects the proper loss function concerning different tasks and goals from the set of loss function candidates.

Given a set of loss function candidates $\{ \ell_{i} \}_{i=1,\cdot \cdot \cdot, n}$ with size $n$, the complexity of search space is $O(n)$. Stochastic loss function (SLF) \cite{Liu_Lai_2020} calculates the overall loss value as follows: 

\begin{equation}
\mathcal{L}(y,\hat{y}) = \sum^{n}_{i=1} \alpha_i \cdot \ell_i(y,\hat{y})
\end{equation}
where a set of weights $\{ \alpha_{i} \}_{i=1,\cdot \cdot \cdot, n}$ represents the contributions of individual loss functions, while $y$ and $\hat{y}$ represent the ground truth and prediction, respectively. However, this soft fusing strategy cannot prohibit the sub-optimal loss function from depreciating the loss value $\mathcal{L}$. To tackle this problem, \textbf{AutoLoss} \cite{zhao2021autoloss} simulates the hard selection with Gumbel-softmax operation on the weight set $\{ \alpha_{i} \}_{i=1,\cdot \cdot \cdot, n}$. Moreover, diverse user-item interaction examples exhibit different convergence behaviors. The weight set for every example cannot be initialized by the same static probability. AutoLoss uses a controller network with several fully-connected layers like equation \ref{equ:MLP}, taking the pair $(y, \hat{y})$ as input and outputting the weight set. Therefore, the weight set $\{ \alpha_{i} \}_{i=1,\cdot \cdot \cdot, n}$ is adaptively produced for different examples depending on distinct convergence behavior patterns.

Although seldom works implement loss function search in recommendation scenario. Automated loss function search itself is not a brand new field direction. In the image semantic segmentation application, AUTO SEG-LOSS \cite{li2020auto} searches particular surrogate losses and enhances the model performance on distinct metrics. In the face recognition task, a search space and a reward-guided search method \cite{wang2020loss} are presented to acquire the best loss function candidate. Those papers validate the effectiveness of the automated loss function search. Perhaps some methodologies can be transferred to recommendation tasks, or new methods can be explicitly proposed for recommendation scenarios. The training efficiency of recommender system models can be further improved.

% Table generated by Excel2LaTeX from sheet 'Sheet1'
% Table generated by Excel2LaTeX from sheet 'Sheet1'
\begin{table*}[htbp]
  \centering
  \caption{Summary of Auto-OCS methods.}
  \begin{adjustbox}{max width=\textwidth}
    \begin{tabular}{ccccc}
    \toprule
    \textbf{Categories} & \textbf{Methods} & \textbf{Search Space} & \textbf{Search Strategies} & \textbf{Tasks} \\
    \midrule
    \midrule
    Loss Function Search & AutoLoss (KDD'2021) \cite{zhao2021autoloss} & $O({3 \choose n})$ & MLP   & CTR \& 5-star ratings  \\
    \midrule
    \multirow{4}[8]{*}{Feature Interaction Function Search} & SIF (WWW'2020) \cite{yao2020efficient} & $O({5 \choose n})$ & Gradient based & Top-K recommendation  \\
\cmidrule{2-5}          & OptInter (ICDE'2022) \cite{lyu2021memorize} & $O(3^{F^2})$ & Gradient based & CTR \\
\cmidrule{2-5}          & AutoFeature (CIKM'2020) \cite{khawar2020autofeature} & $O(5^llF^K)$  & Naive Bayes tree & CTR \\
\cmidrule{2-5}          & AutoPI (SIGIR'2021) \cite{meng2021general} & $ O(6^{n^2})$ & Gradient based & CTR \\
    \bottomrule
    \end{tabular}%
    \end{adjustbox}
  \label{table:autoocs}%
\end{table*}%

\subsection{Feature Interaction Function Search}

The effectiveness of feature interaction has been addressed in recent techniques, and various search methods for beneficial interactions are introduced above. However, most literature uses the same feature interaction function to model all the feature interactions while neglecting their distinction. Wide\&Deep \cite{cheng2016wide} implements a shallow component and MLP to model low- and high-order interactions. Attention networks distinguish the contributions of $2^{nd}$ order interaction in AFM \cite{xiao2017attentional} and DIN \cite{zhou2018deep}. Only simple inner product operations are employed in PNN \cite{qu2016product} and DeepFM \cite{guo2017deepfm}. PIN \cite{zhang2016deep} presents a net-in-net architecture to model the pairwise interactions, accompanied by a product-based neural network. All the literature mentioned above implements the same network architecture or inner product to learn interactions regardless of the input data change, which leads to sub-optimal performance. Therefore, feature interaction search methods are needed to search for suitable interaction functions according to different datasets and tasks as shown in Fig.~\ref{fig:autoocs}. We summarize these methods in Table~\ref{table:autoocs}.

Simple neural interaction functions (\textbf{SIF}) \cite{yao2020efficient} has been proposed to automatically select interaction function for collaborative filtering (CF) \cite{koren2015advances}. Within the search space, interaction function $f$ between user $i$ and item $j$ is designed as $ f(u_i,v_j) = o(g(\mathbf{e}_i), g(\mathbf{e}_j))$. $g(\cdot)$ is the simple non-linear element-wise operation (small MLP with fixed architecture) and $o(\cdot)$ is the vector-wise operations selected from five candidates. Let $n$ denote the number of selected operations. The search space is $O({5 \choose n} )$, including popular human-designed interaction functions. Moreover, inspired by an efficient NAS method based on proximal iterations (NASP) \cite{yao2020efficient1}, SIF implements a one-shot search algorithm jointly searching interaction functions and updating the learning parameters through stochastic gradient descent.

Unlike SIF searches proper functions for all feature interactions. \textbf{OptInter} \cite{lyu2021memorize} divides the potential feature function into three categories for every $2^{nd}$ order feature interaction: element-wise product, MLP, and memorized methods, which regard interactions as a new feature and assign trainable weights. There are three function candidates for each interaction, and $O(F^2)$ $2^{nd}$ order feature interactions exist. Therefore, the complexity of the search space is $O(3^{F^2})$. The element-wise product for $\mathbf{e}_i$ and $\mathbf{e}_j$ is represented as:

\begin{equation}
\mathbf{e}_i \circ \mathbf{e}_j  = [e^1_i \times e^1_j, e^2_i \times e^2_j,\cdot \cdot \cdot,e^s_i \times e^s_j]
\end{equation}

Gradient descent-based NAS search method with Gumbel-softmax operation is implemented to make the discrete search space continuous, and the architecture parameters can be learned by Adam optimizer \cite{kingma2014adam}.

\textbf{AutoPI} \cite{meng2021general} introduces the computational graph to search space. The computational graph is a directed acyclic graph (DAG) comprising input nodes, $n$ intermediate nodes, and the output node in an ordered sequence. $n$ is the pre-defined parameter that controls the complexity of the computational graph. Every node $i$ contains a feature matrix $\mathbf{E}^{(i)} \in \mathbb{R}^{F \times s}$ stacked by $F$ feature embeddings with size $s$, and every directed edge $e_{i,j}$ between node $i$ and node $j$ represents an interaction function $o_{i,j}(\cdot)$, which transforms the feature matrix $\mathbf{E}^{(i)}$. Every intermediate node is calculated using all of its predecessors' values: 
\begin{equation}
\mathbf{X}^{(j)} = \sum^{j-1}_{i=0} o_{i,j}(\mathbf{E}^{(i)})
\end{equation}

There are $\frac{n(n-1)}{2}$ edges, and for each edge, one interaction function is searched from an interaction function set $\mathcal{O}$, including six candidates. Therefore, the search space is $O(6^{n^2})$. 
Six feature interaction functions candidates are Skip-connection, SENET layer \cite{huang2019fibinet}, Self-attention \cite{song2019autoint}, FM \cite{rendle2010factorization}, Single-layer Perceptron, and 1d Convolution. Skip-connection outputs the same feature matrix $\mathbf{e}^\prime= \mathbf{e} \in \mathbb{R}^{F \times s}$ as the input. Single-layer Perceptron uses a linear transformation to transform a flattened feature into a feature matrix $\mathbf{e}^\prime= \mathbf{e} \cdot \boldsymbol{W} \in \mathbb{R}^{F \times s}$, and 1d Convolution outputs a feature matrix $\mathbf{e}^\prime \in \mathbb{R}^{F \times s}$ through $m$ kernel matrices$\{\boldsymbol{C_i}\}_{i =1,\cdot\cdot\cdot,f} \in \mathbb{R}^{F \times 1 \times 1}$.

In the search stage, inspired by DARTS \cite{liu2018darts}, a gradient descent NAS method converts the combinatorial search problem to a bi-level optimization with continuous relaxation. Node-level parameters $\boldsymbol{\alpha} = \{\alpha_{i,j}\}_{i<j}$ show the weights of different interaction functions, where $\alpha_{i,j}$ is a vector with size $|\mathcal{O}|$. Since the value of one node is determined by all its predecessors, edge-level weights $\boldsymbol{\beta}=\{\beta_{i,j}\}_{i<j}$ represent the contributions of the node $i$ to the node $j$, where $\beta_{i,j}$ is a scalar. The bi-level optimization is formulated with $\alpha,\beta$ as the upper-level parameters and other weights as the lower-level parameters. The one-step approximation \cite{liu2018darts} is implemented to tackle the expensive inner optimization problem by approximating the architecture gradient.

\textbf{AutoFeature} \cite{khawar2020autofeature} extends the search of proper interaction function for every interaction to high-order with a distinct DAG sub-network structure. The DAG contains a pre-defined number of operations selected from five interaction functions: pointwise addition, Hadamard product, concatenation, generalized product, and none. The search space is $O(5^llf^K)$, where $l$ represents the number of operations in the sub-network, $f$ represents the number of features, and $K$ represents the interaction order. Due to the immense search space, a tree of Naive Bayes classifiers (NBTree) with thresholds indicating the $90^{th}$ percentile of explored space partitions the search space of architectures recursively. As for the search strategy, the likelihood of selecting from a node is proportionate to the number of samplings formerly selected from that node. 

SIF \cite{yao2020efficient} chooses the proper feature interaction function for all $2^{nd}$ order interactions while neglecting that different feature interactions may require different methodologies to model. Based on SIF, AutoFeature \cite{khawar2020autofeature} searches individual functions for every interaction and extends the order of interactions, including most interaction modeling techniques, but high-order interaction and complex DAG components lead to immense search space. Besides, the search strategy of AutoFeature \cite{khawar2020autofeature} can be trapped in the local optimal. OptInter \cite{lyu2021memorize} balances the efficiency and the generalization of the search space, searching distinct functions for every interaction limited to $2^{nd}$ order, and introduces memorized methods as the candidate of feature interaction functions. Taking feature interactions as new features benefits the model performance since this action makes the correlated patterns of some interactions with strong signals more accessible to be captured. It is worth mentioning that the abuse of memorized methods may depreciate the overall performance due to the overfitting problem accompanied by sparse new features.

\section{Horizontal Comparison for AutoRecSys}
\label{sec:experiment}
The empirical analysis aims to help practitioners investigate the bottlenecks and strengths of current AutoRecSys models from different aspects (e.g., number of parameters, training time, etc.). Then researchers can evaluate the applicability of the existing method to their unique problems or scenarios. Therefore, we select the representative AutoRecSys methods to perform the empirical analyses for two tasks: click-through prediction (CTR) and Top-K recommendation. 

\subsection{Experiment details and hyper-parameter setting}
Following the experiment settings in the original papers, we employ commonly-used metrics, Recall@10 (i.e., recall at rank 10) and NDCG@10 (i.e., normalized discounted cumulative gain at rank 10) for Top-K recommendation, and Area Under the ROC Curve (AUC) for click-through prediction (CTR). To gauge the complexity of the model space, we additionally count the number of model parameters, represented as $\#Params$, and measure the training time in seconds. All the methods are implemented by the codes provided by the authors, and the hyper-parameters are set relying on the authors' suggestions. DeepFM \cite{guo2017deepfm} is set as the base model. Reduce-LR-on-Plateau scheduler and early stopping \cite{zhu2021open} are employed on all methods. For a fair comparison, the same machine with 32G memory and GeForce RTX 2080Ti is utilized for the experiment.

% \subsection{ Evaluation Protocol}

\subsection{Datasets}
For the CTR task, AutoRecSys are horizontally compared on Criteo, and Avazu \cite{qu2022single} by selecting three methods: MDE, PEP, and DeepLight. For Top-K recommendation, AutoFIS, $L_0$-SIGN, and HIRS as representative methods are evaluated on MovieLens 1M \cite{harper2015movielens}, and Book-crossing \cite{ziegler2005improving}. All four datasets are widely used in surveyed papers and related publications \cite{ginart2021mixed,deng2021deeplight,liu2020autofis}. Data pre-processing and dataset splitting strictly follow the setting in \cite{liu2021learnable,su2022detecting}.
\begin{itemize}
\item \textbf{Criteo}: It is a real-world industry dataset for CTR prediction,
which consists of 45 million users' click records on ads
over one month. Each click record contains 13 numerical
feature fields and 26 categorical feature fields. 
\item \textbf{Avazu}: It consists of 40 million users' click records on ads
over 11 days, and each record contains 22 categorical features.
\item \textbf{MovieLens 1M}: It contains users’ ratings on movies. Each
data sample contains a user and a movie with their corresponding
attributes as features.

\item \textbf{Book-crossing}: It contains users' implicit and explicit ratings of books.
Each data sample contains a user and a book with their corresponding
features.

\end{itemize}

\begin{table*}[htbp]
  \centering
%   \tiny
  \caption{Results of representative models on CTR and Top-K Recommendation tasks.}
  \begin{adjustbox}{max width=\textwidth}
% Table generated by Excel2LaTeX from sheet 'Sheet1'
% Table generated by Excel2LaTeX from sheet 'Sheet1'
% Table generated by Excel2LaTeX from sheet 'Sheet1'
% Table generated by Excel2LaTeX from sheet 'Sheet1'
\begin{tabular}{ccccc|cccccc}
\toprule
\multicolumn{5}{c|}{CTR}              & \multicolumn{6}{c}{Top-K Recommendation} \\
\midrule
Datasets & Methods & AUC   & \#Params(M) & Time(s) & Datasets & Methods & R@10  & N@10  & \#Params(M) & Time(s) \\
\midrule
\midrule
\multirow{3}[6]{*}{Criteo} & MDE   & 0.7814 & 85.8481 & 11281 & \multirow{3}[6]{*}{MovieLens 1M} & AutoFIS & 0.8827 & 0.8585 & 1.9805 & 15120 \\
\cmidrule{2-5}\cmidrule{7-11}      & PEP   & 0.7924 & 21.7152 & \textbf{4327} &       & $L_0$-SIGN & 0.9125 & 0.8913 & \textbf{0.2445} & 35380 \\
\cmidrule{2-5}\cmidrule{7-11}      & DeepLight & \textbf{0.8001} & \textbf{18.3606} & 8917  &       & HIRS  & \textbf{0.9542} & \textbf{0.9461} & 1.9747 & \textbf{1920} \\
\midrule
\multirow{3}[6]{*}{Avazu} & MDE   & 0.7712 & 81.6852 & 13310 & \multirow{3}[6]{*}{Book-crossing} & AutoFIS & 0.8751 & 0.8504 & 5.4815 & 10917 \\
\cmidrule{2-5}\cmidrule{7-11}      & PEP   & \textbf{0.7825} & 18.6045 & \textbf{7834} &       & $L_0$-SIGN & 0.9009 & 0.8803 & \textbf{1.6214} & 30641 \\
\cmidrule{2-5}\cmidrule{7-11}      & DeepLight & 0.7818 & \textbf{15.5404} & 9656  &       & HIRS  & \textbf{0.9252} & \textbf{0.9043} & 12.9902 & \textbf{918} \\
\bottomrule
\label{table:result}
\end{tabular}%

    \end{adjustbox}
\end{table*}

\subsection{Results and Analysis}
Results of representative models on CTR and Top-K Recommendation tasks are shown in Table \ref{table:result}. The best result under each metric is shown in bold. We acquire several useful insights from the horizontal comparison for AutoRecSys.

\begin{itemize}
\item For CTR task, the embedding pruning-based AutoEDS methods PEP and Deeplight can outperform the heuristic-based method MDE. The possible explanation is that, unlike heuristic approaches, pruning-based AutoEDS methods estimate the importance of various dimensions at a lower level (i.e., the embedding level) as opposed to a higher level (i.e., the dimension level).

\item For the Top-K recommendation task, GNN-based models $L_0$-SIGN outperform AutoFIS, demonstrating the power of GNNs for interaction modeling. The model that takes into account higher-order feature interactions (i.e., HIRS) beats work that takes into account pairwise interactions. Therefore, incorporating high-order feature interactions helps to improve prediction performance.

\item The results of some methods are not satisfying compared with the reported scores in their papers. They are mostly caused by the fact that various data pre-processing and splitting methods are commonly used, even on the same datasets. This indicates that uniform data splitting and preparation are urgently expected
in order to directly compare the outcomes of different models and save researchers in this community from repeatedly realizing the baselines on their own.
\item Model efficiency is a significant aspect in real-world recommendation tasks, where the industrial recommender systems must be updated often (for example, once per hour) due to the continuous changes in feature distribution \cite{qu2022single}. The training time for each model is usually omitted in the original paper, but plays an essential role in the industry. Excessive training time prevents them from deployment in daily lives. Some models with slow training procedures may be on account of code implementations. Based on the time in Table \ref{table:result}, HIRS achieves decent recommendation performance while retaining the least training time.
\item Memory consumption is a vital metric when practitioners pay attention to memory-efficient recommender systems \cite{joglekar2020neural,shi2020compositional} because there has been a recent spike in the migration of data and models from cloud servers side to edge devices \cite{shi2016edge} to preserve privacy and timeliness. In this circumstance, edge devices (e.g., smartphones) have limited resources. If a little recommendation performance drop is acceptable, models with a small number of parameters (such as $L_0$-SIGN) would be more favored.

\end{itemize}

\section{Future Directions}
\label{sec:future}

\textbf{Feature Cold Start} Although existing methods could efficiently and adaptively assign feature dimensions to features (fields), new features (fields) may be added in real time in practical recommender systems. How to efficiently assign embedding dimensions to these new features is still an open question. For example, IncCTR \cite{wang2020practical} sets a unified embedding dimension for all features and initializes feature embeddings. It is worth mentioning that feature cold start problem is not the unique direction for Auto-EDS, but influences many categories in our AutoRecSys taxonomy. For instance, Auto-FSS should consider not only the data distributions but also the users' interest shifts and newly emerged features since new contents and new labels are created from individuals or companies daily \cite{chen2018sequential}. How to quickly evaluate the new incoming features based on the existing feature selection model still needs more discussions and experiments from the community.
% based on whether a feature is new or not. 
% The For an existing feature, it inherits from the embedding of the existing model as its initialization. Such inheriting transfers the knowledge in historical data to the model that will be trained incrementally. For a new feature, its embedding is randomly initialized as no prior knowledge is available.

\textbf{Long-tail Features} Generally, most existing Auto-EDS methods assume that high-frequency users/items should have a larger embedding dimension than low-frequency users/items. This is because low-frequency users/items have lesser training data. However, PEP \cite{liu2021learnable} illustrates that simply assigning larger dimensions for high-frequency features is sub-optimal. Thus, the dimension assignment for those long-tail users/items is still a challenging problem due to their sparse data.

% \textbf{Multi-modal feature} Most existing automated feature interaction search methods identify valuable interactions between features of users and items. However, there are immense multi-media elements in modern applications. Capturing the interaction between the multi-media data features and the normal features may benefit the performance of the recommender systems.

\textbf{Theory analysis} Most existing AutoML methods for deep recommender systems show competitive recommendation performance and promising results in finding the model's suitable component. However, seldom works provide solid theory analysis to guarantee the effectiveness of the search strategy. MED \cite{ginart2021mixed} provides a close solution based on strict assumptions, which is not practical in a real-world application. $L_0$-SIGN \cite{su2021detecting} ensures the success of valuable interaction search by revealing its relation with the information bottleneck principle \cite{tishby2000information} and spike-and-slab distribution \cite{mitchell1988bayesian}. The gap between the theory and application scenarios should be bridged so that the concrete theory analysis could provide prior knowledge to guide the design of search space and search strategy.

\textbf{AutoML for on-device recommender systems} Most existing AutoML for deep recommender systems focuses on models deployed on a centralized cloud server, which is universal. These centralized deep recommender systems introduce privacy issues when the information of users is shared with the could server and other users. Therefore on-device recommender systems have aroused people's interest, where the recommender systems are deployed on the users' devices rather than the could. There are mainly two challenges. (1) \textbf{Issue of heterology:} They assume all the devices implement the same architectures of the recommender systems or neglect the difference between devices, such as memory size, computation ability, and latency. (2) \textbf{Issue of limited resources:} Unlike the centralized recommender system, implementing millions of parameters in the model, it is impractical to deploy the immense recommender systems on the devices. AutoML for on-device recommender systems automatically designs heterogeneous recommender systems for heterogeneous devices with several restrictions. To deal with one of the limitations: memory budget, RULE \cite{chen2021learning} is proposed to learn the diversified embedding blocks and customize elastic item embeddings for various devices with different memory constraints. AutoML for on-device recommender systems is challenging and distinctive from traditional AutoML or on-device recommender systems, where it emphasizes the automated input components and heterology.

\textbf{AutoML for Various Important Recommendation Tasks.} 
Different recommendation tasks such as social recommendation \cite{yu2021self,8731382}, sequential recommendation \cite{guo2021gcn,10.1145/3477495.3531775}, POI recommendation \cite{long2022decentralized}, and multi-modality recommendations \cite{sun2018multi} have different data inputs. However, the recommender models designed for each specific task have seldom been integrated with automated machine learning to save experts from overloaded model architecture design. For instance, the fusion function selection for multi-modality RecSys is non-trivial and requires domain experts, especially for heterogeneous sources \cite{hoang2022automars}. AutoSTG \cite{pan2021autostg}, which made the first attempt to combine automated neural architecture search with the spatio-temporal graph prediction, could provide helpful insight into AutoML for POI recommendation. Applying AutoML to various recommendation tasks and conquering the unique challenges in each recommendation scenario with specific input data remain open questions.

\textbf{AutoML for GNNs-based Recommendations.}
Many GNN-based recommendation models \cite{LightGCN,guo2021gcn,10.1145/3397271.3401109} have emerged recently due to the fast development of GNNs. Despite the success, many parameters of the graph neural architectures need to be tuned by heavy manual work and domain knowledge. Therefore, some recent works, such as GraphNAS \cite{gao2019graphnas}, and Auto-GNN \cite{zhou2019auto}, integrate AutoML with GNN to automatically design GNN models. However, in recommendation scenarios, the graph construction plays an essential role in the final performance \cite{wu2020graph}. More work should be conducted on choosing different graph construction (e.g., adding edges between two consecutive items \cite{wu2019session}, adjusting the current sequence graph \cite{ma2020memory}) with AutoML. AutoGSR \cite{chen2022autogsr} only searches for the proper graph and layer aggregators, not unique for recommendation tasks, to construct a well-performed model. The search space of AutoML for GNN RecSys should be compact and extensive to cover the handcraft and unknown GNN-based recommendation models.

\section{Conclusion}
\label{sec:conclusion}
Over the past few years, deep recommender systems have become powerful and practical tools for both the academic world and industry applications, and AutoML has emerged as a promising way to automate the design of some components or the entirety of the machine learning pipeline. This survey has conducted a comprehensive review of AutoML methodologies for deep recommender systems and provided a new taxonomy to classify those methods into five categories according to the encountered issues. Finally, we have proposed six potential future research directions.

\bibliographystyle{ACM-Reference-Format}
\bibliography{TOIS-2022-0271}

%%
%% If your work has an appendix, this is the place to put it.
\appendix
\end{document}